\documentclass[preprint,aps,prd,showpacs,nofootinbib]{revtex4}

\usepackage{graphicx}
\usepackage{amsfonts}

\newcommand*{\chpt}{\raise0.4ex\hbox{$\chi$}PT}
\newcommand*{\schpt}{S\raise0.4ex\hbox{$\chi$}PT}
\newcommand*{\ie}{\textit{i.e.},\ }
\newcommand*{\eg}{\textit{e.g.},\ }
\newcommand*{\etc}{\textit{etc.}}

\newcommand*{\et}{\textit{et al.}}

\newcommand*{\npb}[1]{Nucl.\ Phys.\ \textbf{B#1}}
\renewcommand*{\prd}[1]{Phys.\ Rev.\ \textbf{D#1}}

\newcommand*{\npbps}[1]{Nucl.\ Phys.\ \textbf{B}
	(Proc.\ Suppl.) \textbf{#1}}

\newcommand*{\boston}{presented at the International Symposium,
{\it Lattice 2002}, Boston, June 24--29, 2002, to be published
in Nucl.\ Phys.\ {\bf B} (Proc.\ Suppl.)}

\newcommand*{\MeV}{{\rm Me\!V}}

\newcommand*{\Tr}{\textrm{Tr}}
\newcommand*{\tr}{\textrm{tr}}

\newcommand{\cD}{\mathcal{D}}

\newcommand{\cI}{\mathcal{I}}

\newcommand{\cL}{\mathcal{L}}
\newcommand{\cM}{\mathcal{M}}

\newcommand{\cO}{\mathcal{O}}

\newcommand{\cU}{\mathcal{U}}
\newcommand{\cV}{\mathcal{V}}

\def\eq#1{eq.~(\ref{eq:#1})}
\def\eqs#1#2{eqs.~(\ref{eq:#1}) and (\ref{eq:#2})}
\def\eqsthree#1#2#3{eqs.~(\ref{eq:#1}), (\ref{eq:#2}) and (\ref{eq:#3})}

\def\semitimes{\mathrel>\joinrel\mathrel\triangleleft}
\def\ltwid{\raise.3ex\hbox{$<$\kern-.75em\lower1ex\hbox{$\sim$}}}
\begin{document}

\bibliographystyle{apsrev}

\title{Pion and Kaon masses in Staggered Chiral Perturbation Theory}
\author{C. Aubin}
\author{C. Bernard}
\affiliation{Washington University, St. Louis, MO 63130}
\begin{abstract}
We show how to compute chiral logarithms that take into account both the
$\cO(a^2)$ taste-symmetry breaking of staggered fermions and 
the fourth-root trick that produces one taste per flavor.
The calculation starts from the Lee-Sharpe
Lagrangian generalized to multiple flavors.
An error in a previous treatment by one  of us is explained and corrected.
The one loop chiral logarithm corrections to the pion and
kaon masses in the full (unquenched), partially quenched,
and quenched cases are computed as examples.  
\end{abstract}

\maketitle

\section{Introduction}\label{sec:intro}

For simulating fully dynamical lattice QCD at light quark masses,
staggered (Kogut-Susskind, KS) fermions have the advantage of being
very fast relative to other available methods \cite{CHIRAL_PANEL_2001}.  
In addition, an exact chiral symmetry for
massless quarks is retained at finite lattice spacing.  However, 
the advantage in speed of KS fermions may be offset
by systematic issues: on
present realistic lattices (\eg recent MILC simulations
\cite{IMP_SCALING,IMP_SCALING2,MILC_SPECTRUM,FB_LAT02} with $a \approx
0.13\ \textrm{fm}$), the KS taste\footnote{We use the term ``taste''
to describe the staggered symmetry induced by doubling; the taste
symmetry becomes $SU(4)_L\times SU(4)_R$ in the massless, continuum
limit, but is broken at $\cO(a^2)$.  We reserve the term ``flavor''
for true ($u$, $d$ and $s$) flavor.}  violations are not negligible.
Indeed, despite the fact that the MILC simulations use an improved
(``Asqtad'') action that reduces taste violations to $\cO(\alpha_S^2 a^2)$, these
effects can still introduce significant lattice artifacts.

Since one can control the taste of the external particles
explicitly in the simulation, taste-violating artifacts show up
primarily in loop diagrams.  In particular, any quantity or
computation that is sensitive to chiral (pseudoscalar meson) loops can
be expected to show large artifacts at current lattice spacings.  In
order to perform controlled chiral extrapolations and extract physical
results with small discretization errors from staggered simulations,
it is necessary to include the effects of taste violations explicitly
in the chiral perturbation theory (\chpt) calculations to which the
simulations are compared.  The goal of this paper is to develop such a
``staggered chiral perturbation theory'' (\schpt).

One can think of the MILC simulations as introducing flavor with
separate KS fields for $u$, $d$ and $s$ quarks.  The 4 tastes for each
field are then reduced to 1 by taking the fourth root of the quark
determinants for each flavor.\footnote{Since $m_u$ is always chosen
equal to $m_d$ in the MILC simulations, one actually uses a slightly
simpler procedure in practice. Only two KS fields are introduced, and
the square root of the $u,d$ determinant is taken.  However, assuming
algorithmic effects (step-size errors, autocorrelations) are under
control, the two approaches are equivalent. We therefore prefer to
consider the conceptually simpler case where each KS field represents
a single flavor.}  The theory with $\root 4 \of {\rm Det}$ does not
have a local lattice action, and there is some concern that
non-universal behavior may thereby be introduced in the continuum
limit. If we are able to show, by comparing simulations to \schpt\
forms, that the staggered theory produces the expected chiral behavior
in the continuum limit with controlled $\cO(a^2)$ errors, it should go
a long way toward easing worries about the $\root 4 \of {\rm Det}$
trick.

A starting point for any \schpt\ calculation is the work of Lee and
Sharpe \cite{LEE_SHARPE}, who derived the $\cO(a^2)$ chiral Lagrangian
for a single KS field (1 flavor, 4 tastes).  In
Ref.~\cite{CHIRAL_FSB}, a generalization of the Lee--Sharpe Lagrangian
to multiple quark flavors was introduced to calculate chiral loop
effects.  However, there are subtleties in the generalization that
were not appreciated in Ref.~\cite{CHIRAL_FSB}, leading to errors in
the multi-flavor chiral Lagrangian and hence in the final
chiral-logarithm formulas.  These same subtleties also turn out to
have implications even for the tree-level comparison (in
Ref.~\cite{LEE_SHARPE}) of the 1-flavor theory with simulations.

Below we will follow the outlines of the three-step procedure introduced in 
Ref.~\cite{CHIRAL_FSB}, which we restate here for completeness:

\begin{itemize}
\item[1.\ ]{} Generalize the Lee-Sharpe Lagrangian to correspond to
$n$ staggered quark fields, resulting in a (broken) $SU(4n)_L\times
SU(4n)_R$ chiral theory.  Where convenient, we will specialize to the
case of interest, $n=3$.  We call the $n=3$ theory the ``$4\!+\!4\!+\!4$''
theory, since it has three flavors, each with four tastes; its
symmetry is a broken $SU(12)_L\times SU(12)_R$.

\item[2.\ ]{} Calculate one loop quantities (such as $m^2_{\pi_5}$) in the 
$4\!+\!4\!+\!4$ theory.

\item[3.\ ]{} Adjust, by hand, the result to a single taste per flavor
in order to correspond to the physical case (and to simulation data).
This adjustment corresponds to the $\root 4 \of {\rm Det}$ trick.  It
requires an understanding of the correspondence between the meson
diagrams at the chiral level and the underlying quark diagrams and is
basically the ``quark flow'' technique of Ref.~\cite{SHARPE_QCHPT}.
For non-degenerate quark masses, we call the adjusted case the $1\!+\!1\!+\!1$
theory; when we take $m_u=m_d\equiv m_l$ (which corresponds to the
MILC simulations) we call it the $2\!+\!1$ theory.
\end{itemize}

The difficulties in Ref.~\cite{CHIRAL_FSB} arose in step 1. Fierz
transformations were used to simplify the flavor structure in the
taste-symmetry breaking potential.  However, Ref.~\cite{LEE_SHARPE}
had already employed Fierz transformations to simplify the form of
this potential.  The two transformations turn out not to be
compatible. In the Lee-Sharpe case, there was only one flavor, so this
was not an issue. By properly taking into account the mixing of the
flavor indices, we find that two of the six terms in the
symmetry-breaking potential of Ref.~\cite{CHIRAL_FSB} are incorrect.

Another difference with Ref.~\cite{CHIRAL_FSB} is that there $n$ was
taken to be 2, and step 3 was modified to adjust the $u,d$ loops
according to a $\sqrt{\rm Det}$, rather than a $\root 4 \of {\rm Det}$
trick. This was due to the fact that Ref.~\cite{CHIRAL_FSB} took
$m_u=m_d$ from the beginning. However, the entire procedure is much
clearer if every quark flavor is treated equivalently.  Further, we
will see that it is important to be able to treat directly charged
pions (\eg $u\bar d$) that are composed of two independent flavors
transforming under an exact lattice flavor symmetry (when $m_u=m_d$).
Finally, the calculation is actually simpler when we keep all three
quark masses unequal. The fact that the Goldstone charged pion mass
squared must then have an overall factor of $m_u+m_d$ gives a very
useful check on our calculation.

Generalizing the taste-breaking potential properly has lead us to
realize that flavor-neutral mesons in certain taste-nonsinglet channels
can mix at tree-level due to ``hairpin'' diagrams.  We can now
see that such diagrams are present even in one-flavor 
\chpt\ \cite{LEE_SHARPE}; their effects have however not been
appreciated previously.
The coefficients of the hairpin diagrams that arise here are new parameters
in the chiral theory and
have to be fit with simulation data or determined perturbatively.

This paper mirrors the format of Ref.~\cite{CHIRAL_FSB}. In
Sec.~\ref{sec:ls-lag}, we generalize the Lagrangian of Lee and Sharpe,
properly taking into account the flavor and taste structures
involved. Sec.~\ref{sec:pi-K_mass} discusses the calculation of the
one loop chiral logarithms for the flavor-nonsinglet Goldstone meson
mass in the $4\!+\!4\!+\!4$ theory.  It is convenient at this point to
generalize the calculation to the partially quenched case, where the
valence and sea quark masses are completely non-degenerate.  The
results are actually most simply expressed in this case, since there
is a clear distinction between valence and sea quark effects, and no
degeneracies arise that lead to cancellations.  We then make the
transition to the $1\!+\!1\!+\!1$ theory in Sec.~\ref{sec:8+4->2+1}. We write
down results for both the partially quenched and ``full'' 
(equal valence and sea quark masses) cases, focusing primarily
there on features which are different from Ref.~\cite{CHIRAL_FSB}.
The results for the quenched chiral logarithms are discussed in
Sec.~\ref{sec:quenched}. Section~\ref{sec:final_results} adds in the
analytic terms and gives a compendium of final results, in full,
partially quenched, and quenched cases.  In the full 
$m_u=m_d\equiv m_l$ ($2\!+\!1$) case, the results from
Sec.~\ref{sec:final_results} have already been reported in
Ref.~\cite{LAT02}.
We conclude with remarks about other uses for \schpt\ in
Sec.~\ref{sec:conc}.
An Appendix gives some additional details about the symmetries
of the theory and briefly discusses the possible existence of a heretofore
unknown phase of the staggered theory.
This possibility is
however apparently unrealized for physical values of the quark masses.

\section{Generalization of Lee-Sharpe Lagrangian}
\label{sec:ls-lag}

Lee and Sharpe \cite{LEE_SHARPE} describe
pseudo-Goldstone bosons with a non-linearly realized $SU(4)_L\times
SU(4)_R$ symmetry, which originate from a single KS field. This KS
field describes four continuum tastes of quarks.

The $4\times 4$ matrix $\Sigma$ is defined by
\begin{equation}\label{eq:sigma}
	\Sigma \equiv \exp(i\phi/f),
	\quad \phi\equiv\sum_{a=1}^{16} \pi_a T_a 
\end{equation}
where the $\pi_a$ are real, $f$ is the tree-level pion decay constant
(normalized here so that $f_{\pi} \approx 131 \ \MeV$), and the
Hermitian generators $T_a$ are
\begin{equation}\label{eq:T_a}
	T_a = \{ \xi_5, i\xi_{\mu5}, i\xi_{\mu\nu}, \xi_{\mu}, \xi_I\}.
\end{equation}
Here we use the Euclidean gamma matrices $\xi_{\mu}$, with
$\xi_{\mu\nu}\equiv \xi_{\mu}\xi_{\nu}$ ($\mu <\nu$ in
\eq{T_a}), $\xi_{\mu5}\equiv \xi_{\mu}\xi_5$, and $\xi_I \equiv
I$ is the $4\times 4$ identity matrix. The field $\Sigma$ transforms
under $SU(4)_L\times SU(4)_R$ as $\Sigma \rightarrow L\Sigma
R^{\dagger}$.

As discussed in Ref.~\cite{CHIRAL_FSB}, we will keep the singlet meson
$\pi_I\propto \tr\phi$ in this formalism. Due to the anomaly, the
singlet receives a large contribution (which we will call $m_0$) to
its mass, and thus does not play a dynamical role. Lee and Sharpe do
not include this field in their formalism, which is equivalent to
keeping the singlet in and taking $m_0\rightarrow\infty$ at the 
end of the calculation \cite{SHARPE_SHORESH}.  We keep the singlet
here since in the generalized case of $n$ KS fields, it is only the
$SU(4n)$ singlet that is  heavy. In the $m_0\rightarrow\infty$
limit, the other $SU(4)$ singlets will still play a dynamical
role.

The (Euclidean) Lee-Sharpe Lagrangian is then\footnote{Aside 
from the $m_0^2$ term, we need not worry 
about $\pi_I$ dependence in this
Lagrangian, since we are taking the $m_0\rightarrow\infty$ limit. It is
only in the quenched case (Sec.~\ref{sec:quenched}), where we are
unable to take the $m_0\rightarrow\infty$ limit, that we will have 
to examine other $\pi_I$ terms.}
\begin{equation}\label{eq:l-s_lag_4}
	\cL_{(4)} = \frac{f^2}{8} \tr(\partial_{\mu}\Sigma
	\partial_{\mu}\Sigma^{\dagger}) - \frac{1}{4}\mu m f^2
	\tr(\Sigma + \Sigma^{\dagger}) + \frac{2m_0^2}{3}(\pi_I)^2 +
	a^2\cV,
\end{equation}
where $\mu$ is a constant with units of mass, and $\cV$ is the
KS-taste breaking potential.  Correct through $\cO(a^2,m)$ in the dual
expansion in $a^2$ and $m$, we have 
\begin{eqnarray}\label{eq:V_4}
	-\cV \equiv \sum_{k=1}^6 C_k \cO_k & = & C_1
         \tr(\xi_5\Sigma\xi_5\Sigma^{\dagger}) \nonumber \\*
	& & + C_2\frac{1}{2} [ \tr(\Sigma^2) - 
	\tr(\xi_5\Sigma\xi_5\Sigma) + h.c.] \nonumber \\*
	& & +C_3\frac{1}{2} \sum_{\nu}[ \tr(\xi_{\nu}\Sigma
	\xi_{\nu}\Sigma) + h.c.] \nonumber \\*
	& & +C_4\frac{1}{2} \sum_{\nu}[ \tr(\xi_{\nu 5}\Sigma
	\xi_{5\nu}\Sigma) + h.c.] \nonumber \\*
	& & +C_5\frac{1}{2} \sum_{\nu}[ \tr(\xi_{\nu}\Sigma
	\xi_{\nu}\Sigma^{\dagger}) - \tr(\xi_{\nu 5}\Sigma
	\xi_{5\nu}\Sigma^{\dagger})] \nonumber \\*
	& & +C_6\ \sum_{\mu<\nu} \tr(\xi_{\mu\nu}\Sigma
	\xi_{\nu\mu}\Sigma^{\dagger}).
\end{eqnarray}
The 16 pions fall into 5 $SO(4)$ representations with tastes given by
the generators $T_a$. This comes from the ``accidental'' $SO(4)$
symmetry of the potential $\cV$.  We can determine the tree-level
masses of the pions by expanding \eq{l-s_lag_4} to quadratic order:
\begin{equation}\label{eq:tree_lev_mass_4}
	m^2_{\pi_B} = 2\mu m + \frac{4m_0^2}{3}\delta_{B,I}
		+a^2\Delta^{(1)}(\xi_B),
\end{equation}
where $B \in \{5,\mu5,\mu\nu(\mu <\nu),\mu,I \}$. The
$\Delta^{(1)}(\xi_B)$ term comes from the $\cV$ term, and is
given\footnote{In Refs.~\cite{LEE_SHARPE,CHIRAL_FSB}, these
corrections are denoted as $\Delta(\xi_B)$. When we generalize to
multiple KS flavors, we will wish to distinguish this single flavor
$\Delta^{(1)}(\xi_B)$ from the $n$-flavor $\Delta(\xi_B)$.}  in
Refs.~\cite{LEE_SHARPE,CHIRAL_FSB} as:
\begin{eqnarray}\label{eq:delta_4}
	\Delta^{(1)} (\xi_5) & = & 0 \nonumber \\*
	\Delta^{(1)} (\xi_{\mu5}) & = & \frac{16}{f^2}\left( 
	C_1 + C_2 + 3C_3 + C_4 - C_5 + 3C_6 \right) \nonumber \\*
	\Delta^{(1)} (\xi_{\mu\nu}) & = & \frac{16}{f^2}\left( 
	2C_3 + 2C_4 + 4C_6\right) \nonumber \\*
	\Delta^{(1)} (\xi_{\mu}) & = & \frac{16}{f^2}\left( 
	C_1 + C_2 + C_3 + 3C_4 + C_5 + 3C_6 \right) \nonumber \\*
	\Delta^{(1)} (\xi_I) & = & \frac{16}{f^2}\left( 
	4C_3 + 4C_4 \right).
\end{eqnarray}
The vanishing of $\Delta^{(1)} (\xi_5)$ is due to the taste
nonsinglet $U_A (1)$ symmetry
\begin{equation}\label{eq:U_A_sym}
	\Sigma \rightarrow e^{i\theta\xi_5}\Sigma e^{i\theta\xi_5},
\end{equation}
which is unbroken by the lattice regulator, making $\pi_5$ 
a true Goldstone boson.

We now wish to generalize to the case of multiple KS fields. In
Ref.~\cite{CHIRAL_FSB}, for two KS quark fields, this was accomplished
by promoting $\Sigma$ and the mass matrix to $8\times 8$ matrices. In
the general case of $n$ KS fields, which we discuss here, these become
$4n\times 4n$ matrices.  The kinetic energy and mass terms are
correctly given in Ref.~\cite{CHIRAL_FSB}. The only difficulty arises
in generalizing the taste-symmetry breaking potential (or equivalently
the taste matrices $\xi_B$). The generalization of $\cV$ in
Ref.~\cite{CHIRAL_FSB} uses a Fierz transformation on the various
four-quark operators to bring them into a ``flavor unmixed'' form as
follows:
\begin{equation}\label{eq:four-quark}
	\bar{q}_i (\gamma_S\otimes\xi_T)q_i 
	\bar{q}_j (\gamma_{S'}\otimes\xi_{T'})q_j ,
\end{equation}
where $q$ is the quark field,
$i,j$ are $SU(n)$ flavor indices, $\gamma_S$ and $\gamma_{S'}$
are spin matrices, and $\xi_T$ and $\xi_{T'}$ are taste
matrices\footnote{In Ref.~\cite{LEE_SHARPE}, these are referred to as
KS-flavor matrices and denoted by $\xi_F$ and $\xi_{F'}$}. Treating
the taste matrices as spurion fields, we see that for flavor unmixed
4-quark operators,
the $\xi$
are singlets under the flavor
$SU(n)$ symmetry. We can thus make the replacement:
\begin{equation}\label{eq:xi_B}
	\xi_B \rightarrow \xi_B^{(n)} = \left( \begin{array}{cccc}
	     		\xi_B & 0 & 0 & \cdots\\*
     		 	0 & \xi_B & 0 &  \cdots\\*
			0 & 0 & \xi_B &  \cdots \\*
			\vdots &\vdots & \vdots &\ddots\end{array}
	     		\right),
\end{equation}
where the $\xi_B^{(n)}$ are $4n\times 4n$ matrices, and the $\xi_B$ on
the right hand side are still $4\times 4$ taste matrices.

Lee and Sharpe, however, already use Fierz transformations on the
operators in Appendix A of Ref.~\cite{LEE_SHARPE} to ensure that the
final six operators in \eq{V_4} are all single-trace objects. We now
find that the transformation used in Ref.~\cite{CHIRAL_FSB} does not
keep the operators in the same single-trace form.

To see this, let us first assume we have made the replacement~(\ref{eq:xi_B})
in the taste-symmetry breaking potential. The operators $\cO_2$ and $\cO_5$
are then not invariant under axial rotations of the individual fields. For
example, consider a taste $U_A(1)$ transformation on a single
flavor only:
\begin{eqnarray}\label{eq:axial_sym}
	\Sigma &\rightarrow & e^{i\theta\Xi_5}\Sigma 
	e^{i\theta\Xi_5}, 
	\qquad 	\Xi_5 = \left( \begin{array}{cccc}
	     		\xi_5 & 0 & 0 & \cdots\\*
     		 	0 & 1 & 0 &  \cdots\\*
			0 & 0 & 1 &  \cdots \\*
			\vdots &\vdots & \vdots &\ddots\end{array} \right),
\end{eqnarray}
where $\Xi_5$ is a $4n\times 4n$ matrix, shown here as composed of
$4\times 4$ blocks. It is simple to verify that the operators $\cO_1$,
$\cO_3$, $\cO_4$, and $\cO_6$ are invariant under
\eq{axial_sym}. However, using $e^{i\theta\xi_5} = \cos\theta +
i\xi_5\sin\theta$, one finds that $\cO_2$ and $\cO_5$ are not
invariant, and thus are not the correct generalization of the
Lee-Sharpe terms to $n$ flavors.

One approach to generalizing the Lee-Sharpe Lagrangian correctly is
therefore to consider all the different ways that the flavor indices
on the various $\Sigma$ fields in \eq{V_4} can contract. To
do this, we write everything as $4\times 4$ matrices and show the
flavor indices explicitly. For example, the form of $\cO_2$ from
Ref.~\cite{CHIRAL_FSB} can be written as:
\begin{equation}
	\cO^{incorrect}_2 = \frac{1}{2} [ \tr(\Sigma_{ij}\Sigma_{ji}) - 
	\tr(\xi_5\Sigma_{ij}\xi_5\Sigma_{ji}) + h.c.],
\end{equation}
where $\xi_5$ the $4\times 4$ object, and $i$ and $j$ are the
$SU(n)$ flavor indices, to be summed over. Another 
$SU(n)$ invariant we can create with this operator is:
\begin{equation}\label{eq:correct_O2}
	\cO_2 = \frac{1}{2} [ \tr(\Sigma_{ii}\Sigma_{jj}) - 
	\tr(\xi_5\Sigma_{ii}\xi_5\Sigma_{jj}) + h.c.].
\end{equation}
One can easily see that this operator is invariant under
\eq{axial_sym}.

By starting with the other operators in \eq{V_4}, we can similarly
find other correctly generalized terms.  This would for instance alter
$\cO_5$ along the same lines as \eq{correct_O2}.  However, a problem
with this approach is that it is difficult to ensure that the most
general taste-violating potential is generated.  For example, the
operator $\tr(\Sigma_{ii}\Sigma_{jj}^{\dagger} -
\xi_5\Sigma_{ii}\xi_5\Sigma_{jj}^{\dagger})$ is invariant under
\eq{axial_sym} but is not easy to find starting with \eq{V_4}.  That
is because Lee and Sharpe have already set
$\tr(\Sigma\Sigma^{\dagger}) = const.$ in arriving at their $\cO_1$.

A more direct way to find the final form of the taste-breaking
potential involves starting from the quark level and using the
original analysis of Lee and Sharpe instead of their final result.  At
the quark level, gluon exchange can change taste and color, but not
flavor.  Therefore the taste-violating 4-quark operators are composed
of products of two bilinears, each of which is a flavor-singlet, as in
\eq{four-quark}.  The 4-quark operators may be mixed or
unmixed in color.\footnote{In Ref.~\cite{LEE_SHARPE}, color-mixed
operators are Fierzed to put them in a standard, color-unmixed form.
But this is precisely what we do not want to do here because it would
mix the flavor indices.}

To $\cO(a^2)$ in the dual $a^2,m$ expansion, the taste-breaking
operators can be computed in the chiral limit. Since gluon emission
does not change chirality, each bilinear is separately chirally
invariant.  The only such bilinears are vector and axial vector in the
naive theory, which correspond to ``odd'' operators in the staggered
theory (operators in which quark and antiquark fields are separated by
1 or 3 links) \cite{LEPAGE}. Thus only the odd-odd 4-quark operators
in Appendix A of Ref.~\cite{LEE_SHARPE} are relevant to us here. Each
such operator can occur in color mixed and color unmixed form, but
that does not affect the correspondence to \schpt\
operators.\footnote{The color structure does affect the coefficients
of the \schpt\ operators, but since these coefficients are arbitrary
at the chiral level anyway, color mixing is irrelevant here.}  The
even-even operators of Ref.~\cite{LEE_SHARPE} were obtained by
Fierzing the odd-odd operators and may be ignored: They correspond to
flavor-mixed 4-quark operators.

The above reasoning implies that
the arguments in Ref.~\cite{CHIRAL_FSB} were in fact
correct, but only if the replacement
\eq{xi_B} is implemented \textit{before} the Fierz transformations
in Ref.~\cite{LEE_SHARPE} that
put the chiral operators in single-trace form. 
Writing the potential as $\cV = \cU +
\cU\,'$, we then obtain:
\begin{eqnarray}
	\label{eq:U}
	-\cU \equiv \sum_{k} C_k \cO_k & = & C_1
         \Tr(\xi^{(n)}_5\Sigma\xi^{(n)}_5\Sigma^{\dagger}) \nonumber \\*
	& & +C_3\frac{1}{2} \sum_{\nu}[ \Tr(\xi^{(n)}_{\nu}\Sigma
	\xi^{(n)}_{\nu}\Sigma) + h.c.] \nonumber \\*
	& & +C_4\frac{1}{2} \sum_{\nu}[ \Tr(\xi^{(n)}_{\nu 5}\Sigma
	\xi^{(n)}_{5\nu}\Sigma) + h.c.] \nonumber \\*
	& & +C_6\ \sum_{\mu<\nu} \Tr(\xi^{(n)}_{\mu\nu}\Sigma
	\xi^{(n)}_{\nu\mu}\Sigma^{\dagger}) \\*
	\label{eq:U_prime}
	-\cU\,' \equiv \sum_{k'} C_{k'} \cO_{k'} & = & C_{2V}\frac{1}{4} 
		\sum_{\nu}[ \Tr(\xi^{(n)}_{\nu}\Sigma)
	\Tr(\xi^{(n)}_{\nu}\Sigma)  + h.c.] \nonumber \\*
	&&+C_{2A}\frac{1}{4} \sum_{\nu}[ \Tr(\xi^{(n)}_{\nu
         5}\Sigma)\Tr(\xi^{(n)}_{5\nu}\Sigma)  + h.c.] \nonumber \\*
	& & +C_{5V}\frac{1}{2} \sum_{\nu}[ \Tr(\xi^{(n)}_{\nu}\Sigma)
	\Tr(\xi^{(n)}_{\nu}\Sigma^{\dagger})]\nonumber \\*
	& & +C_{5A}\frac{1}{2} \sum_{\nu}[ \Tr(\xi^{(n)}_{\nu5}\Sigma)
	\Tr(\xi^{(n)}_{5\nu}\Sigma^{\dagger}) ],
\end{eqnarray}
where $\Tr$ is the full $4n\times 4n$ trace, and the $\xi^{(n)}_B$ are
$4n\times 4n$ matrices as in \eq{xi_B}. The terms that comprise $\cU$
were found in Ref.~\cite{CHIRAL_FSB}. Now, however, there are no terms
that directly correspond to the operators $\cO_2$ and
$\cO_5$. Instead, we have the four terms in $\cU\,'$.\footnote{The
combination $O_{2V}+O_{2A}$ can be Fierzed into the correct version
of $O_2$, \protect{\eq{correct_O2}}, and similarly for $O_{5A}-O_{5V}$
and the correct version of $O_5$. The other linear combinations are new 
here, but could have been Fierzed into other operators of 
Ref.~\protect{\cite{LEE_SHARPE}} if there were no flavor indices.} It turns out
that only two combinations of the four constants in $\cU\,'$ enter in
the 1-loop result: $C_{2V} - C_{5V}$ and $C_{2A} - C_{5A}$.  The terms
corresponding to $C_{2V} + C_{5V}$ and $C_{2A} + C_{5A}$ do not appear
at this level.

Note that the ``accidental'' $SO(4)$ symmetry of the 
one-flavor theory \cite{LEE_SHARPE}
survives in \eqs{U}{U_prime}, as seen by the fact that the the taste indices
are contracted in a ``Lorentz invariant'' way.  This implies that the
degeneracies of the one-flavor theory will also appear in the $n$-flavor 
case: all four taste-vector pions of a given flavor will be degenerate,
as will all taste-tensors, \textit{etc}.
See the Appendix for further discussion.

For $n$ KS flavors, $\Sigma=\exp(i\Phi / f)$ is a $4n \times 4n$
matrix, and $\Phi$ is given by:
\begin{eqnarray}\label{eq:Phi}
	\Phi = \left( \begin{array}{cccc}
     		 U  & \pi^+ & K^+ & \cdots \\*
     		 \pi^- & D & K^0  & \cdots \\*
		 K^-  & \bar{K^0}  & S  & \cdots \\*
		\vdots & \vdots & \vdots & \ddots \end{array} \right),
\end{eqnarray}
where $U = \sum_{a=1}^{16} U_a T_a$, \etc, with the $T_a$ from
\eq{T_a}. The component fields of the diagonal (flavor-neutral)
elements ($U_a$, $D_a$, \etc) are real; while the other (charged)
fields are complex ($\pi^+_a$, $K^0_a$, \etc), such that $\Phi$ is
Hermitian. Here the $n=3$ portion of $\Phi$ is shown explicitly.  The
mass matrix is now generalized to the $4n\times 4n$ matrix
\begin{eqnarray}
	\cM = \left( \begin{array}{cccc}
     		m_u I  & 0 &0  & \cdots \\*
     		0  & m_d I & 0  & \cdots \\*
		0  & 0  & m_s I  & \cdots\\*
		\vdots & \vdots & \vdots & \ddots \end{array} \right),
\end{eqnarray}
where again, the portion shown is for the $n=3$ case.

Thus, our (Euclidean) Lagrangian becomes:
\begin{eqnarray}\label{eq:final_L}
	\cL & = & \frac{f^2}{8} \Tr(\partial_{\mu}\Sigma 
	\partial_{\mu}\Sigma^{\dagger}) - 
	\frac{1}{4}\mu f^2 \Tr(\cM\Sigma+\cM\Sigma^{\dagger})
	+ \frac{2m_0^2}{3}(U_I + D_I + S_I + \cdots)^2 + a^2 \cV,
\end{eqnarray}
where the $m_0^2$ term includes the $n$ flavor-neutral fields and
$\cV=\cU+\cU\,'$ is given in \eqs{U}{U_prime}.  The $\xi^{(n)}_B$ in
$\cV$ are block-diagonal $4n\times 4n$ matrices, as in \eq{xi_B}.

When the masses vanish, the chiral Lagrangian, \eq{final_L}, has 
a flavor $SU(n)$ vector symmetry and the individual
$U_A(1)$ symmetries for each flavor, both of which were used above,
as well as overall fermion number conservation.
These symmetries actually extend to a $U(n)_\ell \times U(n)_r$
``residual chiral group,'' although this full symmetry is not 
particularly important to us in the present context.  Details
are relegated to the Appendix.

Expanding $\cL$ to quadratic order in meson fields, the potential
$\cU$ gives different masses to different taste mesons, but because
it consists entirely of single-trace terms, the contribution is
independent of the meson flavor.  However, since $\cU\,'$ consists of
two-trace terms, it contributes only to the masses of flavor-neutral
mesons, and in particular only those with vector and axial vector
tastes.  Thus, even at tree-level and with $m_u=m_d$, a $\pi^+$ of a
given taste receives different mass corrections than a neutral $U$ or
$D$ of the same taste.  In simulations, disconnected propagators for
taste-nonsinglet pions (including the Goldstone pion) have invariably
been dropped.  This implies that simulations describe $\pi^+$ mesons,
not those constructed from a single flavor, which would have
disconnected contributions.  The comparison in Ref.~\cite{LEE_SHARPE}
of the 1-flavor \schpt\ tree-level results to simulations is therefore
not justified, although almost all of the conclusions of
Ref.~\cite{LEE_SHARPE} survive a revised treatment.

We thus want a chiral theory with both $u$ and $d$ quarks, even if we
are interested in the $m_u=m_d$ case.  This is the primary reason that
we consider the $4\!+\!4\!+\!4$ theory here rather than the $4\!+\!4$ theory of
Ref.~\cite{CHIRAL_FSB}.\footnote{We remark however that it would in
principle be possible to extract the $\pi^+$ results from a $K^+$
calculation in a (partially quenched) $4\!+\!4$ theory.}

From \eq{final_L}, the tree-level masses of the mesons are:
\begin{equation}\label{eq:tl_masses_12}
	m^2_{M_B} = \mu (m_a + m_b) + a^2\Delta(\xi_B),
\end{equation}
where $a$ and $b$ refer to the two quarks which make up the meson $M$,
and we have defined:
\begin{eqnarray}\label{eq:deltas}
	\Delta (\xi_5) & \equiv & \Delta_P = \Delta^{(1)} (\xi_5) = 0
		\nonumber \\*
	\Delta (\xi_{\mu5}) & \equiv & \Delta_A = \frac{16}{f^2}\left( 
	C_1 + 3C_3 + C_4 + 3C_6 \right) \nonumber \\*
	\Delta (\xi_{\mu\nu})  & \equiv &\Delta_T =
	 	\Delta^{(1)} (\xi_{\mu\nu}) = 
		\frac{16}{f^2}\left(2C_3 + 2C_4 + 4C_6\right) \nonumber \\*
	\Delta (\xi_{\mu}) & \equiv & \Delta_V = \frac{16}{f^2}\left( 
	C_1 + C_3 + 3C_4 + 3C_6 \right) \nonumber \\*
	\Delta (\xi_I)  & \equiv & \Delta_I =
	 \Delta^{(1)} (\xi_I) = \frac{16}{f^2}\left( 
	4C_3 + 4C_4 \right).
\end{eqnarray}
Note that the $m_0^2$ terms and the terms from $\cU\,'$ are not
included in these masses. Those terms, which affect only
flavor-neutral mesons and give non-diagonal contributions in the basis
of \eq{Phi}, will be treated as vertices and summed to all orders
below. Thus the $\cO(a^2)$ corrections in \eq{tl_masses_12}
are flavor independent.

Simulations with the ``Asqtad'' action \cite{MILC_SPECTRUM} give approximately equal splittings
of the mass-squares of various taste mesons in the order  $M_5$, $M_{\mu5}$, 
$M_{\mu\nu}$, $M_\mu$, $M_I$.  From \eq{deltas}, this indicates that
$C_4$ is the dominant coefficient, a conclusion first noted in Ref.~\cite{LEE_SHARPE}.

Upon expanding $\cU\,'$ in \eq{U_prime} to quadratic order,
we find a two-point vertex mixing the taste-vector, flavor-neutral
mesons ($U_{\mu}$, $D_{\mu}$, \etc):
\begin{equation}\label{eq:mix_vertex_V}
	-a^2 \frac{16}{f^2} (C_{2V} - C_{5V})\equiv -a^2 \delta'_V\ .
\end{equation}
In other words, there is a term $+{a^2 \delta'_V\over2} (
U_{\mu}+D_{\mu}+S_{\mu}+\cdots)^2$ in $\cL$.  The vertex in the chiral
theory is shown in Fig.~\ref{fig:2-pt_vertex}(a); while the
corresponding underlying quark diagram is shown in
Fig.~\ref{fig:2-pt_vertex}(b).  There is also a vertex mixing the
taste-axial, flavor-neutrals ($U_{\mu5}$, $D_{\mu5}$, \etc):
\begin{equation}\label{eq:mix_vertex_A}
	-a^2 \frac{16}{f^2} (C_{2A} - C_{5A})\equiv -a^2 \delta'_A\ ,
\end{equation}
\ie a term $+{a^2 \delta'_A\over2} (
U_{\mu5}+D_{\mu5}+S_{\mu5}+\cdots)^2$ in $\cL$.  Similarly, the
$m_0^2$ term in $\cL$ produces a vertex $-4m_0^2/3$ between the
taste-singlet, flavor-neutrals ($U_{I}$, $D_{I}$, \etc).

We thus have to resum the flavor-neutral propagators in three cases:
taste-vector, taste-axial, and taste-singlet.  The methods of Appendix
A in Ref.~\cite{UNPHYSICAL} allow us to calculate the full
flavor-neutral meson propagators easily and write them explicitly in
terms of the true propagator poles (mass eigenstates).  Here we sketch
a few steps in this process. For concreteness we focus explicitly on
the taste-vector case, although the taste-axial case is obtained simply by
replacing $V$ with $A$ in the equations below. The taste-singlet
($m_0^2$) case can be calculated similarly, although a more standard
approach is also possible.  We write the full inverse propagator as:
\begin{equation}\label{eq:full_G}
	G_V^{-1} = G_{0,V}^{-1} + H^V,
\end{equation}
with
\begin{equation}\label{eq:G0}
	(G_{0,V}^{-1})_{MN} = (q^2 + m_{M_V}^2)\delta_{MN},
\end{equation}
\begin{equation}\label{eq:V_cd}
	H^V_{MN} = a^2 \delta'_V\ , \quad  \forall\ M,N.
\end{equation}
Here and below we use $V$ for generic taste-vector states, rather than
the index $\mu$. The indices $M$ and $N$ refer to the flavor-neutral
mesons in the original basis of \eq{Phi}, with $m_{M_V}$ and $m_{N_V}$
the ``unmixed'' masses from \eq{tl_masses_12} (\ie without including
the mixing of \eq{mix_vertex_V}).  For example, in the $n=3$ case,
these mesons are $U_V$, $D_V$, and $S_V$.  Using
Ref.~\cite{UNPHYSICAL}, we then find that:
\begin{eqnarray}\label{eq:prop}
	G_V& =& G_{0,V} + \cD^V  \nonumber\\*
	\cD^V & \equiv &  - G_{0,V} H^V G_{0,V} 
		\frac{\det(G_{0,V}^{-1})}{\det(G_V^{-1})}\ .
\end{eqnarray}
$ \cD^V$ is the part of the taste-vector flavor-neutral propagator
that is disconnected at the quark level (\ie
Fig.~\ref{fig:2-pt_vertex} plus iterations of intermediate sea quark
loops).  We can write this explicitly in terms of the masses as:
\begin{eqnarray}
	\cD^V_{MN}&=& 
	-a^2\delta'_V \frac{1}{(q^2 + m_{M_V}^2)(q^2 + m_{N_V}^2)}\ 
	 \frac{\det(G_{0,V}^{-1})}{\det(G_V^{-1})} \label{eq:D_term1}\\*
	 &&\nonumber\\*  &= & -a^2\delta'_V 
	\frac{{\displaystyle\prod_{L}(q^2 + m_{L_V}^2)}}
	{(q^2 + m_{M_V}^2)(q^2 + m_{N_V}^2)
	{\displaystyle\prod_{F}(q^2 + m_{F_V}^2)}}\ . \label{eq:D_term}
\end{eqnarray}
Here $L$, like $M$ and $N$, labels the unmixed flavor-neutral mesons
in the original basis ($m^2_{L_V}$ are the poles of $G_{0,V}$); while
$F$ indexes the eigenvalues of the full mass matrix ($m^2_{F_V}$ are
poles of $G_V$).  For $n=3$, we name the corresponding full
eigenstates in the taste-vector case $\pi_V^0$, $\eta_V$, and
$\eta'_V$ in analogy with the physical, flavor-neutral, taste-singlet
eigenstates.  We emphasize, however, that all these taste-nonsinglet
particles (including $\eta'_V$ and the corresponding taste-axial
particle $\eta'_{A}$) are physically merely varieties of ``pions:''
pseudoscalars that do not couple to pure-glue states in the continuum
limit, unlike the real $\eta'$.

It is easy to generalize eqs.~(\ref{eq:prop},\ref{eq:D_term}) to
incorporate partial quenching. Iterating Fig.~\ref{fig:2-pt_vertex}(b)
to determine the full propagator generates internal quark loops. Only
sea (unquenched) quarks are therefore allowed in this iteration. Thus,
if the number of sea quarks is $n_S$, the product over $L$ in the
numerator of \eq{D_term} includes only the $n_S$ unmixed
flavor-neutral mesons built from these sea quarks. Likewise, 
only the $n_S$ full eigenvalues are included in the denominator
product over $F$.  The external mesons $M$ and $N$, however, may be
any flavor-neutral states, made from either sea quarks or valence
quarks.  Similarly, in the quenched case $\cD^V_{MN}$ is simply
\begin{equation}\label{eq:D_term_quenched}
	\cD^{V,\rm quench}_{MN}= -a^2\delta'_V 
		\frac{1}{(q^2 + m_{M_V}^2)(q^2 + m_{N_V}^2)}\ .
\end{equation}

Below we will also need the relation
\begin{eqnarray}\label{eq:det_rewrite}
	\frac{\det(G_V^{-1})}{\det(G_{0,V}^{-1})} & = &
	 1 + \tr(G_{0,V} H^V) \nonumber\\*
	& = & 1 + a^2 \delta'_V \sum_{L}
	\Biggl( \frac{1}{q^2 + m_{L_V}^2} \Biggr) \ .
\end{eqnarray}
Here the sum over $L$ is again over the unmixed flavor-neutral mesons
in the original basis.  (In the partially quenched case, only mesons
made from sea quarks are included in the sum.)  This relation allows
one to transform between the result (\ref{eq:D_term}) and the form
\cite{CBMG_PQCHPT} one gets directly by iterating the 2-point vertex,
\eq{mix_vertex_V}.

Equations~(\ref{eq:full_G}) through (\ref{eq:det_rewrite}) apply
explicitly to the taste-vector case; to get the taste-axial case, just
let $V\to A$.  These formulas can also be used for the taste-singlet
($I$) channel with the replacement $a^2\delta'_V\rightarrow 4m^2_0/3$.
We get:
\begin{equation}\label{eq:D_term_I}
	\cD^I_{MN}= 
-\frac{ 4m^2_0}{3}\  \frac{{\displaystyle\prod_{L}(p^2 + m_{L_I}^2)}}
		{(p^2 + m_{M_I}^2)(p^2 + m_{N_I}^2)
{\displaystyle{\prod}_{F}(p^2 + m_{F_I}^2)}}\qquad \ ,
\end{equation}
where $L$ and $F$ have the same meaning as in \eq{D_term}.
The $m_0^2\to \infty$ limit in the $4\!+\!4\!+\!4$ case is easily obtained, if
desired, using $m^2_{\eta'_I} \cong 4m^2_0$ for large $m_0^2$. The
$\eta'_I$ then decouples.  However, we prefer not to take the
$m_0^2\to \infty$ limit at this stage, because the form of the
result is then
slightly different in the $4\!+\!4\!+\!4$ and $1\!+\!1\!+\!1$ cases, as we will
discuss in Sec.~\ref{sec:8+4->2+1}.

In the quenched case, the product over sea quark states in the
numerator and denominator of \eq{D_term_I} are omitted.  Of course,
$m^2_0$ now cannot be taken to infinity, and the $\eta'_I$ does not
decouple.  It is therefore necessary to consider possible additional
$\eta'_I$ dependent terms in our Lagrangian.  As discussed in
Refs.~\cite{CHIRAL_FSB} and~\cite{CBMG_QCHPT}, one can do this simply
by making the the replacement $m_0^2\rightarrow m_0^2+\alpha q^2$,
where $\alpha$ is an additional quenched chiral parameter.  This gives
\begin{equation}\label{eq:D_term_quenched_I}
	\cD^{I,\rm quench}_{MN}= -\frac{4}{3}\ \frac{m^2_0 + \alpha q^2}
{(q^2 + m_{M_I}^2)(q^2 + m_{N_I}^2)}\ .
\end{equation}

It is sometimes useful to think of the quenched case as the limit of
the partially quenched case as the sea quark masses go to infinity (at
fixed valence masses and fixed $m_0^2$, $\alpha$, $\delta'_V$ and
$\delta'_A$).  For disconnected propagators, the resulting decoupling
of the sea quarks has the simple effect of canceling the ``unmixed''
terms in the numerator with the terms involving the full masses in the
denominator.  Thus \eq{D_term} becomes \eq{D_term_quenched}, and
\eq{D_term_I} becomes \eq{D_term_quenched_I}. (The $\alpha q^2$ term
in \eq{D_term_quenched_I} could have been put in for free in
\eq{D_term_I} since it is irrelevant in the $m_0^2\to \infty$ limit.)

\section{One loop Pion mass for $4\!+\!4\!+\!4$ dynamical flavors}
\label{sec:pi-K_mass}

\begin{figure}
 \includegraphics[width=2in]{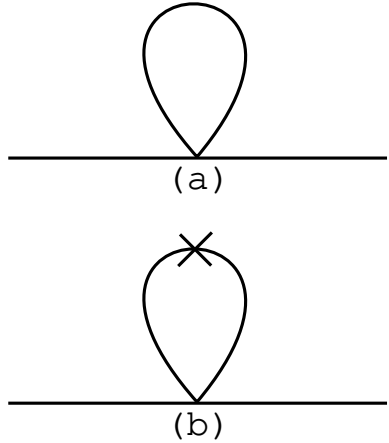} \caption{The only diagrams
	contributing to the flavor-nonsinglet meson self energy. (a) includes
	all contributions where the propagator in the loop contains no
	two-point vertex insertions. (b) subsumes the graphs which
	have disconnected insertions on the propagator. The cross
	represents one or more insertions of either the $m_0^2$, $\delta'_V$,
	or $\delta'_A$ vertices.}
	\label{fig:tadpole}
\end{figure}

We can now calculate the 1-loop Goldstone pion self energy. We shall
use the term ``pion'' to refer to a generic flavor-nonsinglet meson
here, so it can refer to the kaons as well (and also what we will
shortly call a $P^+$ meson). As in Ref.~\cite{CHIRAL_FSB}, all the
contributing diagrams are tadpoles, as shown in
Fig.~(\ref{fig:tadpole}), coming from each of the terms in
\eq{final_L}. We can break up the self energy (defined to be {\it
minus}\/ the sum of self energy diagrams) as
\begin{equation}\label{eq:self energy}
	\Sigma(p^2) = \frac{1}{16\pi^2 f^2}\left[ \sigma^{con}(p^2)
	+ \sigma^{disc}(p^2) \right] \ ,
\end{equation}
where ``$con$'' and ``$disc$'' are short for connected and
disconnected, respectively.  The main difference here from
Ref.~\cite{CHIRAL_FSB} is that the disconnected piece (for $m_u\ne
m_d$) now receives contributions from all the terms in the Lagrangian,
not just the mass term. Also, note that we have factored out
$1/16\pi^2 f^2$, and not $1/96\pi^2 f^2$ as in Ref.~\cite{CHIRAL_FSB}.

\begin{figure}
 \includegraphics[width=2in]{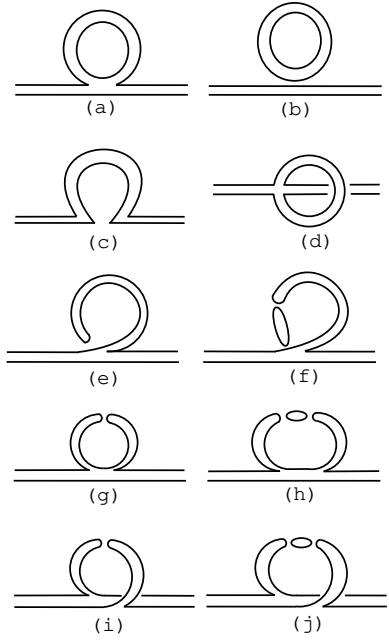} \caption{The quark level
  	diagrams that could contribute to the 1-loop
  	meson self energy. Diagrams (b) and (c), which require vertices of the
	form of Fig.~\protect{\ref{fig:vertices}}(c), do not occur in the case
	of interest.  Similarly, (d) only contributes to flavor-neutral
	propagators.  Note that (f), (h) and (j) correspond to (e), (g), 
	and (i), respectively,
  	with iteration of either $m_0^2$, $\delta'_V$, or $\delta'_A$
  	vertices. These diagrams are to
  	be taken as including any number of iterations, thus multiple
  	internal quark loops.}  \label{fig:diagrams}
\end{figure}

The terms ``connected'' and ``disconnected'' refer to the internal
loop at the quark level.  In other words, a disconnected diagram will
have either an internal disconnected propagator
(Figs.~\ref{fig:diagrams}(g)-(j)) or a disconnected vertex
(Fig.~\ref{fig:diagrams}(e)), or both (Fig.~\ref{fig:diagrams}(f)).
The disconnected propagators correspond to one or more insertions of a
two-point $\delta'$ or $m^2_0$ vertex (\ie $\cD$ in
\eqsthree{D_term}{D_term_quenched}{D_term_I}); while the disconnected
vertices are generated by the $\cU\,'$ term in the potential, as we
will see below.

We will explicitly perform the partially quenched calculation. Here
the quenched valence quarks (call them $x$ and $y$) will in general
have different masses from the sea quarks $u$, $d$ and $s$.  We will
still refer to this as a ``$4\!+\!4\!+\!4$'' partially quenched theory, based
on its dynamical quark content.  The chiral Lagrangian needed has 5
flavors, but 2 flavors are dropped, by fiat, from loops.  From this
5-flavor, partially quenched theory, we can find equivalent
3-flavor (what we call ``full'' theory) results by setting the valence
quark masses equal to various sea quark masses.

The valence quarks $x$ and $y$ form new mesons in our
theory, which we name as follows:
\begin{eqnarray}\label{eq:XYP}
	X & = & x\bar{x} \qquad	Y  =  y\bar{y}\nonumber\\*
	P^+ & = & x\bar{y} \qquad P^-  =  y\bar{x}\ .
\end{eqnarray}
We will not give individual names to the mesons formed from various
valence-sea combinations such as $x\bar{u}$, but just refer to them
generically by ``$Q$.''  A check on our final calculation here is that
the 1-loop correction to $m_{P^+_5}^2$ should be proportional to $m_x
+ m_y$ (and hence $m_{P^+_5}^2$ itself), due to the separate $U_A(1)$
symmetries for $x$ and $y$ and the $x\leftrightarrow y$ interchange
symmetry.

\begin{figure}
 \includegraphics[width=2in]{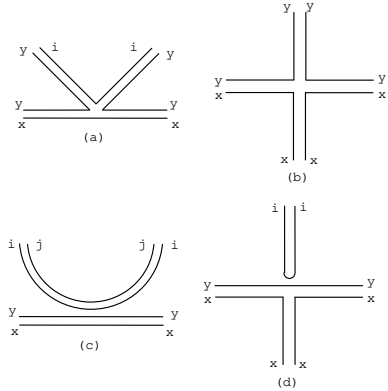} \caption{The possible quark
  	level diagrams for $2\rightarrow 2$ meson scattering, where
  	one incoming and one outgoing particle (shown horizontally)
  	are fixed to be valence mesons, $P^+ =x\bar y$.  The indices
  	$i$ and $j$ represent arbitrary quark flavors.  There are two
  	additional diagrams (not shown), which are like (a) and (d)
  	but have the roles of $x$ and $y$ interchanged.}
  	\label{fig:vertices}
\end{figure}

Since the mass, kinetic, and $\cU$ terms are composed entirely of
single traces, the relevant 4-meson vertices that they generate are
all of the form of Figs.~\ref{fig:vertices}(a) and (b). (This is
because ``touching'' flavor indices must be the same in a single
trace.)  In Fig.~\ref{fig:vertices}(b) the vertical meson lines must
join to make the internal loop; however, they can only join with a
disconnected propagator because they have different flavors.  Thus
connected contributions from mass, kinetic, and $\cU$ all involve the
vertex of Fig.~\ref{fig:vertices}(a), and produce diagrams of the form
of Fig.~\ref{fig:diagrams}(a).  Disconnected contributions can come
from Figs.~\ref{fig:vertices}(a) and (b).

The $\cU\,'$ terms, on the other hand, involve two traces, and
therefore generate disconnected vertices, which in principle can be of
the form of either Fig.~\ref{fig:vertices}(c) or (d).  However it is
not hard to show, from the explicit taste structure of $\cU\,'$, that
only vertices with odd number of mesons coming from each trace
contribute when two of the mesons are Goldstone particles
(pseudoscalar taste). The $\cU\,'$ vertices then separate into two
disconnected pieces, one with a single meson and the other with three.
Thus $\cU\,'$ vertices must be of the form of
Fig.~\ref{fig:vertices}(d), and not (c).  This in turn implies that
the $\cU\,'$ self energy diagrams have the disconnected structure of
Fig.~\ref{fig:diagrams}(e) or (f) only, where (e) uses a connected
propagator and (f), a disconnected one.

Combining the connected contributions, we find:
\begin{eqnarray}\label{eq:connected}
	\sigma^{con}(p^2)&= &
	-\frac{1}{12}\sum_{Q,B}\int \frac{d^4 q}{\pi^2} 
	 \biggl[ p^2 + q^2 
	+\left( m^2_{P^+_5} + m^2_{Q_5} \right)+
	a^2\Delta(\xi_B)
	  \biggr]\frac{1}{q^2 + m^2_{Q_B}}.
\end{eqnarray}
As before, $B$ takes on the taste values \{5, $\mu5$, $\mu\nu \
(\mu<\nu),\ \mu,\ I$\}, and $Q$ runs over all meson flavors with one
valence quark ($x$ or $y$) and one sea quark ($u$, $d$, or $s$).
Which mesons contribute is clear from
Fig.~\ref{fig:diagrams}(a).\footnote{If we were considering a full
$n=5$ flavor theory where $x$ and $y$ were unquenched, then the quark
loop in Fig.~\protect{\ref{fig:diagrams}(a)} could also be an $x$ or
$y$, and the sum over $Q$ would include the mesons $X$, $Y$ and $P$
(\eq{XYP}).}  The first two terms in \eq{connected} come from the
kinetic energy: one from the derivatives acting on the external legs
and the other from the derivatives acting on the internal loop. The
last two terms are from the mass term and $\cU$ respectively. We have
used the fact that $\mu (m_x + m_y) = m_{P^+_5}^2$ to rewrite the
mass-term contribution.

The one-loop mass renormalization is just the self energy with the
external momentum $p^2$ evaluated at $-m_{P^+_5}^2$.  Making this
substitution and noting from \eq{tl_masses_12} that $m^2_{Q_B} =
m^2_{Q_5} + a^2\Delta(\xi_B)$, the term inside the square brackets
becomes $q^2 + m^2_{Q_B}$, which cancels the denominator from the
propagator. Thus, no chiral logarithms arise from these terms.  This
corresponds to the fact that all diagrams of the form of
Fig.~\ref{fig:diagrams}(a) cancel in the standard continuum chiral
logarithm calculation \cite{GASSER_LEUTWYLER}. (See
Ref.~\cite{CHIRAL_FSB} for more discussion.)

For the disconnected contributions, it will be convenient to divide up
$\sigma^{disc}$ further, according to (1) whether the particle in the
loop is a vector, axial-vector, or singlet in taste, and (2) the type of
diagram that generates the term.  We thus have
\begin{equation}\label{eq:sigma_disc}
	\sigma^{disc} = \sigma^{disc}_V + \sigma^{disc}_A
	 + \sigma^{disc}_I \ ,
\end{equation}
with 
\begin{eqnarray}\label{eq:sigma_disc_bydiagram}
	\sigma_V^{disc} &=& \sigma_{V,gh} + \sigma_{V,ij}
	 + \sigma_{V,e} + \sigma_{V,f}\nonumber\\*
	\sigma_A^{disc} &=& \sigma_{A,gh} + \sigma_{A,ij}
	 + \sigma_{A,e} + \sigma_{A,f}\nonumber\\*
	\sigma_I^{disc} &=& \sigma_{I,gh} + \sigma_{I,ij} \ .
\end{eqnarray}
Here, the labels $gh$, $ij$, $e$, and $f$ refer to the diagrams in
Fig.~\ref{fig:diagrams} that generate the contribution.  As discussed
above, the $gh$ and $ij$ contributions come from kinetic energy, mass,
or $\cU$ vertices, with a disconnected propagator.  The $e$ and $f$
contributions have a $\cU\,'$ vertex and a connected or disconnected
propagator, respectively.  As is easily seen from the form
\eq{U_prime}, $\cU\,'$ vertices that have two Goldstone mesons on the
external lines must have only taste-vector or axial mesons on the
loop.  Therefore, $\sigma_I^{disc}$ gets contributions from
Fig.~\ref{fig:diagrams}(g)--(j) only, as it does in the continuum.

We focus first on the taste-vector contributions.  $\sigma_{V,e}$ uses
the vertex Fig.~\ref{fig:vertices}(d) with $i=x$ (or $i$=$y$ in its
$x\to y$ variant) in order to have a connected propagator. We find,
therefore,
\begin{eqnarray}\label{eq:sig_ve}
	\sigma_{V,e} & = & 
	-\frac{2}{3}a^2\delta'_V \int\frac{d^4 q}{\pi^2} \biggl[
	\frac{1}{q^2+m^2_{X_V}} + \frac{1}{q^2+m^2_{Y_V}}
	\biggr] \ ,
\end{eqnarray}
where all but the overall coefficient follows immediately from the
form of the diagram.  We have already included the factor of 4 for the
four degenerate taste-vector mesons, and will continue to do so below.

$\sigma_{V,f}$ again uses the vertex Fig.~\ref{fig:vertices}(d), but
now $i$ must be one of the sea quarks, since a virtual quark loop is
involved.  The propagator is the disconnected taste-vector
propagator, $\cD^V$, \eq{D_term}. We have
\begin{equation}\label{eq:sig_vf}
	\sigma_{V,f}  = -\frac{2}{3} a^2\delta'_V\int 
	\frac{d^4 q}{\pi^2}
	\sum_{M=U,D,S}\left( \cD^V_{XM}+ \cD^V_{YM}\right)
	\ .
\end{equation}
Note that both $\sigma_{V,e}$ and $\sigma_{V,f}$ have explicit factors
of $\delta'_V$.  These come from the 4-meson vertex, generated by
$\cU'$.  There are additional implicit factors of $\delta'_V$ in the
disconnected propagator $\cD^V$ in $\sigma_{V,f}$. It is not
immediately obvious that this same linear combination of $C_{2V}$ and
$C_{5V}$ (see \eq{mix_vertex_V}) must occur in both the 2-meson and
4-mesons vertices.  However, we will see below that it is necessary
for the cancellations that allow the ${P^+_5}$ mass renormalization to
be proportional to $m^2_{P^+_5}$, as required by axial symmetry.

$\sigma_{V,gh}$ is generated by vertices of type Fig.~\ref{fig:vertices}(a),
with $i=y$ (or $i$=$x$ in its $y\to x$ variant).  The result is
\begin{eqnarray}
	\sigma_{V,gh}(p^2)  &= &-\frac{1}{3} \int 
	\frac{d^4 q}{\pi^2}\Biggl[\left(p^2 + q^2 + m^2_{P^+_5} +m^2_{X_5} +a^2\Delta_V
                       \right)\cD^V_{XX} \nonumber\\*
	&&\qquad +\left(p^2 + q^2 + m^2_{P^+_5} + m^2_{Y_5} + a^2 \Delta_V\right) 
                    \cD^V_{YY}\Biggr] \ .
\end{eqnarray}
The $p^2 + q^2$ terms come from the kinetic energy vertex;
$m^2_{P^+_5}$, $m^2_{X_5}$ and $m^2_{Y_5}$, from the mass vertex; and
the $\Delta_V$ terms, from $\cU$.  Putting $p^2 = -m^2_{P^+_5}$, and
using $m^2_{X_V} = m^2_{X_5} + a^2\Delta_V$, from \eq{tl_masses_12},
this simplifies to:
\begin{equation}\label{eq:sig_vgh}
	\sigma_{V,gh}(-m^2_{P^+_5})  = -\frac{1}{3} \int 
	\frac{d^4 q}{\pi^2}\Biggl[\left(q^2 + m^2_{X_V}\right)\cD^V_{XX} \
	 +\left(q^2 + m^2_{Y_V}\right)\cD^V_{YY} \Biggr] \ .
\end{equation}

Finally, we have $\sigma_{V,ij}$. This contribution uses the vertex
Fig.~\ref{fig:vertices}(b) and, clearly, an $X$-$Y$ disconnected
propagator.
\begin{eqnarray}\label{eq:sig_vij}
	\sigma_{V,ij}(p^2) & = &-\frac{2}{3} \int 
	\frac{d^4 q}{\pi^2}\Biggl[\left(p^2 + q^2 - m^2_{P^+_5} +  a^2 \Delta_V\right)\cD^V_{XY} \Biggr] \nonumber\\* 
	\sigma_{V,ij}(-m^2_{P^+_5}) & = &-\frac{2}{3} \int 
	\frac{d^4 q}{\pi^2}\Biggl[\left(q^2 - 2m^2_{P^+_5} +  a^2 \Delta_V\right)\cD^V_{XY} \Biggr]  \ .
\end{eqnarray}

The sum of all the contributions to $\sigma^{disc}_V$ can be
simplified with an identity derived by combining
\eqs{D_term1}{det_rewrite}:
\begin{equation}\label{eq:identity}
	\sum_{M=U,D,S}\cD^V_{XM} =- \frac{1}{q^2 + m^2_{X_V}} 
	- \frac{q^2 + m^2_{X_V}}{a^2\delta'_V}\cD^V_{XX} \ .
\end{equation}
Using this and the $X\leftrightarrow Y$ version we can add \eq{sig_vf} to
\eqs{sig_ve}{sig_vgh}.  The trivial identity (from \eq{D_term})
\begin{equation}\label{eq:trivial}
	(q^2 + m^2_{X_V})\cD^V_{XX} = (q^2 + m^2_{Y_V})\cD^V_{XY} 
\end{equation}
and the fact that $m^2_{X_V}+m^2_{Y_V} = 2m^2_{P^+_V} = 2m^2_{P^+_5} +
2a^2\Delta_V$, can then be used to combine the result with \eq{sig_vij}
to give simply
\begin{equation}\label{eq:sigma_disc_v}
\sigma^{disc}_V (-m^2_{P^+_5})  = 2m^2_{P^+_5} \int \frac{d^4 q}{\pi^2}
	\cD^V_{XY} \ .
\end{equation}
Note that the result is proportional to $m^2_{P^+_5}$, as expected.
The corresponding expression for $\sigma^{disc}_A$ is obtained by
$V\to A$.

For $\sigma^{disc}_I$, we just have contributions from
Figs.~\ref{fig:diagrams}(g)--(j).  These contributions are
very similar to the corresponding ones for $\sigma^{disc}_V$.  We
have:
\begin{eqnarray}
\sigma_{I,gh}(p^2)& = & -\frac{1}{12} \int 
	\frac{d^4 q}{\pi^2}\Biggl[\left(p^2 + q^2 + m^2_{P^+_5} +m^2_{X_5} +a^2\Delta_I
                       \right)\cD^I_{XX} \nonumber\\*
	&&\ \qquad\qquad +\left(p^2 + q^2 + m^2_{P^+_5} + m^2_{Y_5} + a^2 \Delta_I\right) 
                    \cD^I_{YY}\Biggr] \nonumber\\*
\sigma_{I,gh}(-m^2_{P^+_5})& = & -\frac{1}{12} \int 
        \frac{d^4 q}{\pi^2} \left[\left(q^2 +  m^2_{X_I}\right) \cD^I_{XX}
        + \left(q^2 +  m^2_{Y_I}\right) \cD^I_{YY}\right] \label{eq:sig_Igh} \\
\sigma_{I,ij}(p^2)& = & \frac{1}{6} \int 
	\frac{d^4 q}{\pi^2}\left[\left(p^2 + q^2 - m^2_{P^+_5}  +a^2\Delta_I
                       \right)\right]\cD^I_{XY} \nonumber\\*
\sigma_{I,ij}(-m^2_{P^+_5})& = & \frac{1}{6} \int
        \frac{d^4 q}{\pi^2}\left[\left(q^2 - 2m^2_{P^+_5}  +a^2\Delta_I
                       \right)\right]\cD^I_{XY} \label{eq:sig_Iij} \ , 
\end{eqnarray}
The $V\to I$ version of \eq{trivial} allows us to combine
\eqs{sig_Igh}{sig_Iij}, yielding
\begin{equation}\label{eq:sigma_disc_I}
\sigma^{disc}_I (-m^2_{P^+_5}) = -\frac{m^2_{P^+_5}}{2} \int
        \frac{d^4 q}{\pi^2} \cD^I_{XY} 
\end{equation}
Again, the result is proportional to $m^2_{P^+_5}$.

Collecting \eqs{sigma_disc_v}{sigma_disc_I} according to
\eq{sigma_disc}, gives
\begin{equation}\label{eq:sigma_disc_combined}
	\sigma^{disc}(-m^2_{P^+_5}) =  \frac{m^2_{P^+_5}}{2} \int
        \frac{d^4 q}{\pi^2} \left(4 \cD^V_{XY}+4 \cD^{A}_{XY} 
	-\cD^I_{XY}\right) \ .
\end{equation}
Since $\sigma^{con}$ is just a quarticly divergent constant, the above
result
contains all the 1-loop chiral logarithms in the mass renormalization.

The result in \eq{sigma_disc_combined} is rather implicit.  To express
the chiral logarithms more concretely, we would need three further
steps:
\begin{itemize}
\item[]{(1)} find the explicit expressions
for the eigenvalues of the full mass matrices in the denominators of the $\cD$
(\eg $m^2_{\pi^0_V}$, $m^2_{\eta_V}$ and $m^2_{\eta'_V}$ in $\cD^V$).

\item[]{(2)} take the $m_0^2\to \infty$ limit in the taste-singlet
term.  

\item[]{(3)} write the disconnected propagators $\cD$ as sums of
simple poles and perform the integrals over $q$.
\end{itemize}
Steps (1) and (2) are slightly different in the $1\!+\!1\!+\!1$ case of
interest than in the present $4\!+\!4\!+\!4$ case, so we postpone them until
later.  On the other hand, step (3) can be done quite generally, so we
present it here.

The integrands in \eq{sigma_disc_combined} are of the form
\begin{equation}\label{eq:integrand}
\cI^{[n,k]}\left(\left\{m\right\}\!;\!\left\{\mu\right\}\right) 
\equiv \frac{\prod_{a=1}^k (q^2 + \mu^2_a)}
{\prod_{j=1}^n (q^2 + m^2_j)} \ ,
\end{equation}
where $\{m\}$ and $\{\mu\}$ are the sets of masses
$\{m_1,m_2,\dots,m_n\}$ and $\{\mu_1,\mu_2,\dots,\mu_k\}$.
respectively.  As long as there are no mass degeneracies in the
denominator, and $n>k$ (which is true here even after the $m_0^2\to
\infty$ limit), $\cI{[n,k]}$ can be written as the sum of simple poles
times their residues:
\begin{equation}\label{eq:lagrange}
	\cI^{[n,k]}\left(\left\{m\right\}\!;\!\left\{\mu\right\}\right) = 
	\sum_{j=1}^n \frac{R_j^{[n,k]}\left(\left\{m\right\}\!;\!\left\{\mu\right\}\right)}{q^2 + m^2_j}\ ,
\end{equation}
where
\begin{equation}\label{eq:residues}
	R_j^{[n,k]}\left(\left\{m\right\}\!;\!\left\{\mu\right\}\right)
	 \equiv  \frac{\prod_{a=1}^k (\mu^2_a- m^2_j)}
	{\prod_{i\not=j} (m^2_i - m^2_j)}\ .
\end{equation}
Equation (\ref{eq:lagrange}) just follows from the fact that an analytic
function is determined by its poles and behavior at infinity; it is
known as ``Lagrange's formula'' in complex analysis \cite{LAGRANGE}.

The integrals of the simple poles can now be done using
\begin{equation}\label{eq:I1}
	\cI_1 \equiv \int \frac{d^4 q}{(2\pi)^4}
	 \frac{1}{q^2 + m^2} \rightarrow \frac{1}{16\pi^2} 
	\ell( m^2)\ , 
\end{equation}
where
\begin{equation}\label{eq:chiral_log_infinitev}
	\ell(m^2) \equiv  m^2 \ln \frac{m^2}{\Lambda^2}
	\qquad{\rm [infinite\ volume]} \ ,
\end{equation}
with $\Lambda$ the chiral scale. We use the arrow in \eq{I1}
and later to indicate that we are only keeping the chiral logarithm
terms.  If the system is in a finite (but large) spatial volume $L^3$,
we only have to modify \eq{chiral_log_infinitev}:
\begin{equation}\label{eq:chiral_log}
	\ell( m^2) \equiv  m^2 \left(\ln \frac{m^2}{\Lambda^2}
	 + \delta_1(mL)\right) \qquad{\rm [finite\ spatial\ volume]} \ ,
\end{equation}
where \cite{CHIRAL_FSB}
\begin{equation}\label{eq:delta1}
	\delta_1(mL)  =  \frac{4}{mL}
		\sum_{\vec r\ne 0}
		\frac{K_1(|\vec r|mL)}{|\vec r|} \ ,
\end{equation}
with $K_1$ the Bessel function of imaginary argument.

With the above, we can write a general integral of the form in
\eq{sigma_disc_combined} as
\begin{equation}\label{eq:general_integral}
	\int \frac{d^4 q}{\pi^2} 
	\cI^{[n,k]}\left(\left\{m\right\}\!;\!\left\{\mu\right\}\right)
	 \rightarrow
	\sum_{j=1}^n {R_j^{[n,k]}\left(\left\{m\right\}\!;
	\!\left\{\mu\right\}\right)} \ 
	\ell(m_j^2) \ .
\end{equation}

We make one final comment on the $4\!+\!4\!+\!4$ calculation before going on
to the $1\!+\!1\!+\!1$ case. In Ref.~\cite{CHIRAL_FSB}, certain chiral
logarithm terms were claimed to come from pure valence diagrams, with
connected propagators; while in the current calculation, all such
terms cancel.  What is the reason for the discrepancy?  As discussed
above, the problem in \cite{CHIRAL_FSB} was the incorrect treatment of
flavor indices.  Because of this, it was not realized that there is a
difference between a propagator of a flavor-neutral Goldstone pion,
such as $U_5$, and that of the flavor-nonsinglet $\pi^+_5$ (or between their
partially quenched counterparts, $X_5$ and $P^+_5$). An explicit
computation in the current framework shows that the connected, valence
terms found in \cite{CHIRAL_FSB} do in fact exist, but only for a
flavor-neutral propagator.  Such terms arise identically in the
$X_5$-$Y_5$, $X_5$-$X_5$, and
$Y_5$-$Y_5$ propagators, but do not appear in the $P^+_5$-$P^-_5$
propagator.  This proves that they come from
Fig.~\ref{fig:diagrams}(d). The needed vertex is
Fig.~\ref{fig:vertices}(b) (after relabeling), which is generated by
kinetic, mass and $\cU$ terms.  The claim in \cite{CHIRAL_FSB} that
connected, valence terms come from Fig.~\ref{fig:diagrams}(c) with
vertex Fig.~\ref{fig:vertices}(c) is incorrect.  Indeed, it was argued
above that the flavor structure of the terms in our Lagrangian forbids
vertex Fig.~\ref{fig:vertices}(c), at least with two external
Goldstone mesons.

\section{Moving from $4\!+\!4\!+\!4$ to $1\!+\!1\!+\!1$ dynamical flavors}\label{sec:8+4->2+1}

To make the $4\!+\!4\!+\!4$ result, \eq{sigma_disc_combined}, into a $1\!+\!1\!+\!1$
result, we simply must divide by a factor of 4 for every sea quark
loop.  The contributing diagrams are Figs.~\ref{fig:diagrams}
(e)--(j). There can be either taste-vector, taste-axial vector or
taste-singlet mesons on the internal lines of these diagrams, and we
can treat all these cases at the same time simply by defining
\begin{equation}\label{eq:dp_def}
	\delta' = \cases { a^2 \delta'_V, &taste-vector;\cr
	a^2 \delta'_A, &taste-axial;\cr
	4m_0^2/3, &taste-singlet.\cr }
\end{equation}
Diagrams (e), (g), and (i) have no sea quark loops and a single factor
of $\delta'$ (in (e) this comes from the 4-meson $\cU\,'$ vertex).
Diagrams (f), (h), and (j) have one additional factor of $\delta'$ for
each sea quark loop. Therefore, dividing by 4 for every sea quark loop
is the same as dividing every factor of $\delta'$, except the first,
by 4.  For a general function $f(\delta')$ which vanishes linearly as
$\delta'\rightarrow 0$, we can make this adjustment simply by the
replacement $f(\delta') \rightarrow 4
f\left(\frac{\delta'}{4}\right)$.  Alternatively, we can see from
\eqsthree{sigma_disc_combined}{D_term}{D_term_I} that the first factor
of $\delta'$ comes from the explicit $\delta'$ in front of $\cD^V$,
$\cD^{A}$ or $\cD^I$; while higher order terms in $\delta'$ are
implicit in the values of the ``full'' masses in the denominators
relative to the ``unmixed'' masses in the numerators. Therefore, to go
from $4\!+\!4\!+\!4$ to $1\!+\!1\!+\!1$, we leave the explicit $\delta'$ factors alone
but merely let $\delta'\to \delta'/4$ before diagonalizing the full
mass matrix.

The full mass matrices to be diagonalized follow from the
flavor-neutral mixing term in $\cL$, written down following
\eqs{mix_vertex_V}{mix_vertex_A}.  After $\delta'\to \delta'/4$, these
have the form
\begin{equation}\label{eq:mass_matrix}
	\left(\matrix{m^2_U +\delta'/4 & \delta'/4 & \delta'/4 \cr
	\delta'/4 & m^2_D +\delta'/4 & \delta'/4 \cr
	\delta'/4 & \delta'/4 & m^2_S +\delta'/4 \cr}\right)
\end{equation}
Here the masses $m^2_U$, $m^2_D$, $m^2_S$ have an implicit taste label
($V$, $A$, or $I$) depending on which case we are considering.  The
explicit expressions for the eigenvalues of \eq{mass_matrix} are
complicated and not illuminating in general.  The solutions in the
$2\!+\!1$ ($m_u=m_d$) case, however, have simple forms, and
that is the case of greatest current interest.  In the
taste-vector channel, we have, for the $2\!+\!1$ case,
\begin{eqnarray}\label{eq:eigenvalues_V}
        m_{\pi^0_V}^2 & = & m_{U_V}^2 = m_{D_V}^2 \ ,   \nonumber \\*
        m_{\eta_V}^2 & = & \frac{1}{2}\left( m_{U_V}^2 + m_{S_V}^2 +
        \frac{3}{4}a^2\delta'_V - Z
        \right)\ , \nonumber \\*
        m_{\eta'_V}^2 & = &  \frac{1}{2}\left( m_{U_V}^2 + m_{S_V}^2 +
        \frac{3}{4}a^2\delta'_V +  Z \right) \ ; \\
	 Z & \equiv &\sqrt{\left(m_{S_V}^2-m_{U_V}^2\right)^2
	 - \frac{a^2\delta'_V}{2} 
        \left(m_{S_V}^2-m_{U_V}^2\right) +\frac{9(a^2\delta'_V)^2}{16}
	 } \nonumber
\end{eqnarray}
The taste-axial case just requires $V\to A$.  In the taste-singlet
case, $\delta' = 4m_0^2/3$, and $m_0^2$ will be taken to infinity, so
only the large-$m_0$ expressions are needed.  We have (again for
$2\!+\!1$):
\begin{eqnarray}\label{eq:eigenvalues_I}
        m_{\pi^0_I}^2 & = & m_{U_I}^2 = m^2_{D_I}   \nonumber \\*
        m_{\eta_I}^2 & = & \frac{m_{U_I}^2}{3}+
        \frac{2m_{S_I}^2}{3} \\  
        m_{\eta'_I}^2 & = & m_0^2 \ ,\nonumber
\end{eqnarray}
where we have neglected corrections that are $\cO(1/m_0^2)$ compared
to the terms kept.

Finally, we can give the result for the chiral logs in the Goldstone
pion self energy.  For the moment we stay with the partially quenched
expression and also assume no degeneracies among the valence and sea quark
masses.  In the $1\!+\!1\!+\!1$ case we obtain from \eq{sigma_disc_combined}
with \eqsthree{D_term}{D_term_I}{self energy}:
\begin{eqnarray}\label{eq:sigma_111}
	\Sigma^{1+1+1}(-m^2_{P_5^+}) & \rightarrow &
	 \frac{m^2_{P_5^+}}{16\pi^2f^2}\Big( 
	-2a^2\delta'_V \sum_{j_V} R^{[5,3]}_{j_V}\; \ell(m^2_{j_V})
	-2a^2\delta'_A \sum_{j_A} R^{[5,3]}_{j_A}\; \ell(m^2_{j_A})
	 \nonumber \\*
	&&\qquad\qquad+\frac{2}{3} \sum_{j_I} R^{[4,3]}_{j_I}\;
	 \ell(m^2_{j_I})\Big) \ ,
\end{eqnarray}
where we have used \eq{general_integral}, and $R_j^{n,k}$ and
$\ell(m^2)$ are given by eqs.~(\ref{eq:residues}) and
(\ref{eq:chiral_log_infinitev}) or (\ref{eq:chiral_log}).  $j_V$ runs
over $\{{X_V},{Y_V},{\pi^0_V},{\eta_V},{\eta'_V}\}$ and
\begin{equation}\label{eq:R53V_explicit}
	R^{[5,3]}_{j_V} =  R^{[5,3]}_{j_V}\left(\{m_{X_V},
	m_{Y_V},m_{\pi^0_V},m_{\eta_V},m_{\eta'_V}\};
	\{m_{U_V},m_{D_V},m_{S_V}\}\right)\ .
\end{equation}
(For $R^{[5,3]}_{j_A}$, just let $V\to A$). Similarly, $j_I$ runs over
$\{{X_I},{Y_I},{\pi^0_I},{\eta_I}\}$ (the $\eta'_I$ has decoupled in
the $m_0^2\to\infty$ limit), and
\begin{equation}\label{eq:R43I_explicit}
	 R^{[4,3]}_{j_I} =  R^{[4,3]}_{j_I}
	\left(\{m_{X_I},m_{Y_I},m_{\pi^0_I},m_{\eta_I}\};
	\{m_{U_I},m_{D_I},m_{S_I}\}\right)\ .
\end{equation}
In each taste channel, the values of $m^2_{\pi^0}$, $m^2_\eta$, and
$m^2_{\eta'}$ in \eq{sigma_111} are just the eigenvalues of the
corresponding version of \eq{mass_matrix}.

The $2\!+\!1$ ($m_u=m_d$) case is very similar, but because $m_{\pi^0}^2 =
m_{U}^2 = m_{D}^2$, there is a cancellation in \eqs{D_term}{D_term_I}
between $q^2 + m_{\pi^0}^2$ in the denominators and, say, $q^2 +
m_{D}^2$ in the numerators.  Assuming no other degeneracies, we have
\begin{eqnarray}\label{eq:sigma_21}
	\Sigma^{2+1}(-m^2_{P_5^+}) & \rightarrow & 
	\frac{m^2_{P_5^+}}{16\pi^2f^2}\Big( 
	-2a^2\delta'_V \sum_{j_V} R^{[4,2]}_{j_V}\; \ell(m^2_{j_V})
	-2a^2\delta'_A \sum_{j_A} R^{[4,2]}_{j_A}\; \ell(m^2_{j_A})
	 \nonumber \\*
	&&\qquad\qquad+\frac{2}{3} \sum_{j_I} R^{[3,2]}_{j_I}\;
	 \ell(m^2_{j_I})\Big) \ .
\end{eqnarray}
Here $j_V$ runs over $\{{X_V},{Y_V},{\eta_V},{\eta'_V}\}$ and
\begin{equation}\label{eq:R42V_explicit}
	R^{[4,2]}_{j_V} =  R^{[4,2]}_{j_V}\left(
	\{m_{X_V},m_{Y_V},m_{\eta_V},m_{\eta'_V}\};
	\{m_{U_V},m_{S_V}\}\right)\ .
\end{equation}
Again, let $V\to A$ for $R^{[4,2]}_{j_A}$. The index $j_I$ runs over
$\{{X_I},{Y_I},{\eta_I}\}$, and
\begin{equation}\label{eq:R32I_explicit}
	R^{[3,2]}_{j_I} =  R^{[3,2]}_{j_I}\left(
	\{m_{X_I},m_{Y_I},m_{\eta_I}\};
	\{m_{U_I},m_{S_I}\}\right)\ .
\end{equation}
In this case, the values of $m^2_{\pi^0}$, $m^2_\eta$, and
$m^2_{\eta'}$ are given by \eqs{eigenvalues_V}{eigenvalues_I}.

Cases of interest with further degeneracies (such as a ``full'' $2\!+\!1$
pion with $m_x=m_y=m_u=m_d$) can be obtained by carefully taking
limits in \eq{sigma_21}. We will write down some of these cases
explicitly in Sec.~\ref{sec:final_results}, where we also include the
analytic contributions.

\section{Quenched Case}\label{sec:quenched}

Since we can think of the quenched theory as the limit of the 
partially quenched theory as the sea quark 
masses go to infinity, all the
manipulations that led to \eq{sigma_disc_combined} will go through
unscathed in the quenched case.  We can therefore simply replace the
disconnected propagators in \eq{sigma_disc_combined} with their
quenched versions, \eqs{D_term_quenched}{D_term_quenched_I}.  Using
the same notation as in \eqs{sigma_111}{sigma_21}, we have (assuming
$m_x \not= m_y$):
\begin{eqnarray}\label{eq:sigma_quenched}
	\Sigma^{\rm quench}(-m^2_{P_5^+}) & \rightarrow &
	 \frac{m^2_{P_5^+}}{16\pi^2f^2}\Big( 
	-2a^2\delta'_V \sum_{j_V} R^{[2,0]}_{j_V}\; \ell(m^2_{j_V})
	-2a^2\delta'_A \sum_{j_A} R^{[2,0]}_{j_A}\; \ell(m^2_{j_A})
	 \nonumber \\*
	&&\qquad\qquad+\frac{2}{3} \sum_{j_I} R^{[2,0]}_{j_I}\;
	 (m_0^2-\alpha m^2_{j_I}) \ell(m^2_{j_I})\Big) \ .
\end{eqnarray}
Here $j_V$ runs over $\{{X_V},{Y_V}\}$; similarly for $j_A$ and $j_I$.
For the $\alpha$-dependent terms,  we have used the integral
\begin{equation}\label{eq:I2}
	\cI_2 \equiv \int \frac{d^4 q}{(2\pi)^4}
	 \frac{q^2}{q^2 + m^2} = -m^2\cI_1 
	+ \int \frac{d^4 q}{(2\pi)^4}
	  \rightarrow -\frac{1}{16\pi^2} 
	m^2 \ell(m^2) \ ,
\end{equation}
where $\cI_1$ is defined in \eq{I1}.

Because the quenched residues here are particularly simple, it is
useful to write out the result more explicitly:
\begin{eqnarray}\label{eq:Sig_qu}
	\Sigma^{\rm quench}(-m^2_{P_5^+})  
	& \rightarrow & \frac{m^2_{P_5^+}}{16\pi^2f^2} \Biggl[ 
	-2a^2\delta'_V\frac{\ell(m^2_{X_V})
	-\ell(m^2_{Y_V})}{m^2_{Y_V}-m^2_{X_V}}
	-2a^2\delta'_A\frac{\ell(m^2_{X_A})
	-\ell(m^2_{Y_A})}{m^2_{Y_A}-m^2_{X_A}}
	\nonumber\\*
	&& + \frac{2}{3}\frac{(m_0^2-\alpha m^2_{X_I})\ell(m^2_{X_I})  -
	(m_0^2-\alpha m^2_{Y_I}) \ell(m^2_{Y_I})}{m^2_{Y_I}-m^2_{X_I}}
	\Biggr].
\end{eqnarray}

\section{Final one-loop results}
\label{sec:final_results}

The mass at one loop is given by
\begin{equation}\label{eq:pi_m_sq}
	(m_{P^+_5}^{1-{\rm loop}})^2 = m_{P^+_5}^2 + 
	\Sigma(-m_{P^+_5}^2) \ .
\end{equation}
The chiral logarithm contributions to $\Sigma(-m_{P^+_5}^2)$ are
presented in \eqsthree{sigma_111}{sigma_21}{Sig_qu}, but for complete
one-loop expressions we also need the ``$\cO(p^4)$'' analytic terms.
The latter are unchanged from Ref.~\cite{CHIRAL_FSB}.  However, for
the analytic coefficients we now prefer to use the more standard
\cite{GASSER_LEUTWYLER} $L_i$, rather than the parameters $K_3$ and
$K'_4$ employed in \cite{CHIRAL_FSB}.

In the absence of any degeneracies, we have, in the $1\!+\!1\!+\!1$ case,
\begin{eqnarray}\label{eq:mass_an_terms}
	\frac{(m^{1-{\rm loop},\, 1+1+1}_{P^+_5})^2}
	{\left( m_x+m_y \right)}&=& \mu \Biggl\{1 +
	\frac{1}{16\pi^2f^2}\Big( 
	-2a^2\delta'_V \sum_{j_V} R^{[5,3]}_{j_V}\; \ell(m^2_{j_V}) 
	-2a^2\delta'_A \sum_{j_A} R^{[5,3]}_{j_A}\; \ell(m^2_{j_A})
	 \nonumber \\*
	&&\qquad\qquad+\frac{2}{3} \sum_{j_I} R^{[4,3]}_{j_I}\;
	 \ell(m^2_{j_I})\Big) 
	+ \frac{16\mu}{f^2}\left(2L_8-L_5\right)
	\left(m_x+m_y\right) \nonumber\\*
	&&\qquad\qquad +\frac{32\mu}{f^2}\left(2L_6-L_4\right)
	 \left(m_u+m_d+m_s \right) + a^2 C \Biggr\}.
\end{eqnarray}
Definitions here are the same as in \eq{sigma_111}; the chiral
logarithm function $\ell(m^2)$ is given by \eq{chiral_log_infinitev},
or in finite volume, by \eq{chiral_log}.  Recall that ours is a joint
expansion in the quark masses (generically $m$) and $a^2$.  The
analytic terms in $(m^{1-{\rm loop}})^2$ here are $\cO(m^2)$ or
$\cO(ma^2)$; $\cO(a^4)$ terms cannot enter here because the
pion mass must vanish in the chiral limit. 
Lattice effects violating continuum rotational invariance 
cannot show up at this order for the Goldstone pion --- see Appendix.

Similarly, for $m_u=m_d\equiv m_l$ (the $2\!+\!1$ case), but with no other
degeneracies, we have
\begin{eqnarray}\label{eq:mass_an_terms_21}
	\frac{(m^{1-{\rm loop},\, 2+1}_{P^+_5})^2}
	{\left( m_x+m_y \right)}&=& \mu \Biggl\{1 +
	\frac{1}{16\pi^2f^2}\Big( 
	-2a^2\delta'_V \sum_{j_V} R^{[4,2]}_{j_V}\; \ell(m^2_{j_V}) 
	-2a^2\delta'_A \sum_{j_A} R^{[4,2]}_{j_A}\; \ell(m^2_{j_A})
	 \nonumber \\*
	&&\qquad\qquad+\frac{2}{3} \sum_{j_I} R^{[3,2]}_{j_I}\;
	 \ell(m^2_{j_I})\Big) 
	+\frac{16\mu}{f^2}\left(2L_8-L_5\right)
	\left(m_x+m_y\right) \nonumber \\*
	&&\qquad\qquad +\frac{32\mu}{f^2}\left(2L_6-L_4\right)
	 \left(2m_l+m_s \right) + a^2 C \Biggr\}.
\end{eqnarray}
Definitions here are the same as in \eq{sigma_21}.  

The quenched result is
\begin{eqnarray}\label{eq:qu_an_terms}
	\frac{(m^{1-{\rm loop},\, {\rm quench}}_{P^+_5})^2}
	{\left( m_x+m_y \right)}&=&
	 \mu \Biggl\{1 + \frac{1}{16\pi^2 f^2} \Biggl[ 
	-2a^2\delta'_V\frac{\ell(m^2_{X_V})
	-\ell(m^2_{Y_V})}{m^2_{Y_V}-m^2_{X_V}}
	-2a^2\delta'_A\frac{\ell(m^2_{X_A})
	-\ell(m^2_{Y_A})}{m^2_{Y_A}-m^2_{X_A}}
	\nonumber\\*
	&&\qquad\qquad + \frac{2}{3}\frac{(m_0^2-\alpha 
	m^2_{X_I})\ell(m^2_{X_I})  -
	 (m_0^2-\alpha m^2_{Y_I}) \ell(m^2_{Y_I})}{m^2_{Y_I}-m^2_{X_I}}
	\Biggr] \nonumber\\*
	&&\qquad\qquad  + \frac{16\mu}{f^2}\left(2L'_8-L'_5\right)
	\left(m_x+m_y\right) + a^2 C'  \Biggr\},
\end{eqnarray}
where the primes on $L'_8$, $L'_5$, and $C'$ indicate that they may
have different values than in the unquenched cases. Of course, there is
no analytic term involving the sea quarks ($2L'_6-L'_4$) in the
quenched case.

It is useful to write down more explicit versions of the above results
in various limits pertinent to many simulations.  First, with
$m_u\ne m_d$, we set $m_x=m_u$ and $m_y=m_d$ to obtain the ``full
QCD'' charged pion mass in the $1\!+\!1\!+\!1$ case:
\begin{eqnarray}\label{eq:pion_final_nd}
	\frac{(m^{1-{\rm loop},\, 1+1+1}_{\pi^+_5})^2}{\left( m_u+m_d \right)}
	\!\!\!&&=  \mu\Biggl\{1 + \frac{1}{16\pi^2 f^2}\Biggl[
	-2a^2\delta'_V  \Biggl( \frac{m_{S_V}^2 - m^2_{\pi^0_V}}
	{(m_{\eta_V}^2 - m^2_{\pi^0_V})(m_{\eta'_V}^2 -
	m^2_{\pi^0_V})} \ell(m^2_{\pi^0_V})
	+ \nonumber\\*&& \frac{m_{S_V}^2 - m^2_{\eta_V}}
	{(m_{\eta'_V}^2 - m^2_{\eta_V})(m_{\pi^0_V}^2 -
	m^2_{\eta_V})} \ell(m^2_{\eta_V}) + 
	 \frac{m_{S_V}^2 - m^2_{\eta'_V}}
	{(m_{\eta_V}^2 - m^2_{\eta'_V})(m_{\pi^0_V}^2 -
	m^2_{\eta'_V})}  \ell(m^2_{\eta'_V})
	\Biggr)  \nonumber\\*
	&& +\biggl( V\to A \biggr)+ \frac{2}{3}\Biggl(\frac{m_{S_I}^2 
	- m_{\pi^0_I}^2}{m_{\eta_I}^2 - m_{\pi^0_I}^2}
	\ell(m_{\pi^0_I}^2)
	+ \frac{m_{S_I}^2 - m_{\eta_I}^2}
	{m_{\pi^0_I}^2 - m_{\eta_I}^2}\ell(m_{\eta_I}^2) \Biggr)\Biggr]
	\nonumber\\*
	&&\!\!\!\!\!\!\!\!\!\! +\frac{16\mu}{f^2}
	\left(2L_8-L_5\right)\left(m_u+m_d\right) 
	+\frac{32\mu}{f^2}\left(2L_6-L_4\right)
	\left(m_u+m_d+m_s \right) + a^2 C \Biggr\}.
\end{eqnarray}
This result is most easily obtained by taking the degenerate mass
limits in \eq{sigma_disc_combined}, before the integral is performed,
rather than in \eq{mass_an_terms}.  The quantities $m^2_{\pi^0}$,
$m^2_\eta$ and $m^2_{\eta'}$ are eigenvalues of the mass matrix,
\eq{mass_matrix}.  From \eq{pion_final_nd} we can get the charged kaon
mass simply by interchanging the explicit labels $d\leftrightarrow s$ and
$D\leftrightarrow S$. (The neutral labels $\pi^0$, $\eta$, and $\eta'$ 
are unaffected.)

The results for the full pion and kaon in the case of degenerate up
and down quark masses (both set to $m_l$) are also of interest, as
they are needed to fit many simulations. Since the pion and kaon
results look quite different, we show them both:
\begin{eqnarray}\label{eq:pi_K_final_deg}
	\frac{(m^{1-{\rm loop},\, 2+1}_{\pi^+_5})^2}{2m_l}
	&=& \mu\Biggl\{1 + \frac{1}{16\pi^2 f^2}\Biggl(
	 \biggl[ -4\ell(m^2_{\pi^0_V})
	- \frac{2a^2\delta'_V}{m_{\eta'_V}^2 - m^2_{\eta_V}}\biggl(
	 \frac{m_{S_V}^2 - m^2_{\eta_V}}
	{m_{\pi^0_V}^2 -
	m^2_{\eta_V}} \ell(m^2_{\eta_V}) \nonumber\\*
	&&-  \frac{m_{S_V}^2 - m^2_{\eta'_V}}
	{m_{\pi^0_V}^2 -m^2_{\eta'_V}}  \ell(m^2_{\eta'_V})
	\biggr)\biggr]+\biggl[ V\to A \biggr]+ 
	\ell(m_{\pi^0_I}^2)
	 - \frac{1}{3}\ell(m_{\eta_I}^2)\Biggr)\nonumber\\*
	&&+\frac{16\mu}{f^2}\left(2L_8-L_5\right)\left(2m_l \right)
	+\frac{32\mu}{f^2}\left(2L_6-L_4\right)
	 \left(2m_l+m_s \right) + a^2 C \Biggr\},\\*
	\frac{(m^{1-{\rm loop},\, 2+1}_{K^+_5})^2}{(m_l+m_s)}
	&=& \mu\Biggl\{1 + \frac{1}{16\pi^2 f^2}\Biggl(
	\biggl[-\frac{2a^2\delta'_V}{m_{\eta'_V}^2 - m^2_{\eta_V}}
	\biggl( \ell( m^2_{\eta_V})
	- \ell(m^2_{\eta'_V})\biggr)\biggr]
	+\biggl[ V\to A \biggr]\nonumber\\*
	&&\qquad\qquad\qquad+ \frac{2}{3}\ell(m_{\eta_I}^2)\Biggr)
	+\frac{16\mu}{f^2}\left(2L_8-L_5\right)
	\left(m_l+m_s\right) \nonumber\\*
	&&\qquad\qquad\qquad\qquad+\frac{32\mu}{f^2}
	\left(2L_6-L_4\right) \left(2m_l+m_s \right) + a^2 C \Biggr\}.
\end{eqnarray}
Again, the relevant limits are most easily taken before the integrals
are performed. The $\pi^0$, $\eta$ and $\eta'$ masses in this case are
given explicitly in \eqs{eigenvalues_V}{eigenvalues_I}; we have made
heavy use of these explicit forms to simplify the chiral logarithm
terms in the $\pi$ mass.

The last case we will look at is the quenched pion mass correction in
the limit of degenerate valence masses ($m_y=m_x$). Here we
get a double pole in the pion self energy.  We can either carefully
take the limit $m_y\to m_x$ in \eq{qu_an_terms}, or return to
\eq{sigma_disc_combined} with quenched $\cD$ terms and do the double
pole integrals directly. We follow the latter approach.  We need the
following integrals:
\begin{eqnarray} \label{eq:Integral3}
	\cI_3 &\equiv& \int \frac{d^4 q}{(2\pi)^4}
	 \frac{1}{(q^2 + m^2)^2} = -\frac{\partial}
	{\partial m^2}\cI_1\rightarrow
	\frac{1}{16\pi^2}  \tilde \ell(m^2) \ , \\
	\label{eq:Integral4}
	\cI_4 &\equiv& \int \frac{d^4 q}{(2\pi)^4}
	 \frac{q^2}{(q^2 + m^2)^2} = \cI_1-m^2\cI_3
	\rightarrow
	\frac{1}{16\pi^2}  \left(\ell(m^2) - m^2\tilde \ell(m^2)\right) \ ,
\end{eqnarray}
where $\cI_1$ is given in \eq{I1}; $\ell(m^2)$, in
\eq{chiral_log_infinitev}; and
\begin{equation}\label{eq:elltilde_infinitev}
	\tilde \ell(m^2) \equiv  -\left(\ln 
	\frac{m^2}{\Lambda^2} + 1\right)\qquad
	{\rm [infinite\ volume]} \ ,
\end{equation}
with $\Lambda$ the chiral scale.  In finite spatial volume $L^3$,
\begin{equation}\label{eq:elltilde}
	 \tilde \ell(m^2) \equiv  -\left(\ln 
	\frac{m^2}{\Lambda^2} + 1\right) + \delta_3(mL) 
	\qquad{\rm [finite\ spatial\ volume]} \ ,
\end{equation}
where \cite{CHIRAL_FSB}
\begin{equation}\label{eq:delta3}
	\delta_3(mL)  = 2 \sum_{\vec r\ne 0}
		K_0(|\vec r|mL)\ ,
\end{equation}
with $K_0$ the Bessel function of imaginary argument.  Note that the
$+1$ term in $\tilde\ell(m^2)$ was omitted in Ref.~\cite{CHIRAL_FSB}.
That is formally acceptable when we are only keeping chiral
logarithms, but inconvenient, since then the result from performing
the double pole integral is not equal to degenerate limit of the chiral
logs from the single poles.

Using the above integrals, we get:
\begin{eqnarray}\label{eq:Sig_qu_deg}
	\frac{(m^{1-{\rm loop},\, {\rm quench}}_{P^+_5})^2}{2m_x}
	&= & \mu \Biggl\{1 + \frac{1}{16\pi^2 f^2}\Biggl(
	-2a^2\delta'_V  \tilde \ell(m^2_{X_V}) 
	-2a^2\delta'_A  \tilde \ell(m^2_{X_A}) 
	+  \frac{2m_0^2}{3} \tilde \ell(m^2_{X_I})
	\nonumber \\* 
	&&\!\!\!\!+ \frac{2\alpha}{3} 
	 (\ell(m^2_{X_I}) -m^2_{X_I}\tilde \ell(m^2_{X_I}) \Biggr) 
	+ \frac{16\mu}{f^2}\left(2L'_8-L'_5\right)
	\left(2m_x\right) + a^2 C'  \Biggr\},
\end{eqnarray}
Taking the $m_y\to m_x$ limit in \eq{qu_an_terms} of course gives the
same result.  To see that the finite-size corrections are the same
both ways, one needs the identity \cite{CHIRAL_FSB}
\begin{equation}
	\delta_3(mL)  =   -\delta_1(mL)-{mL\over2}\delta'_1(mL) \ .
\end{equation}

Double poles also appear in some other interesting limits of
\eqs{mass_an_terms}{mass_an_terms_21}.  For example, the ``partially
quenched degenerate pion'' in either the $2\!+\!1$ case
($m_x=m_y\not=m_l$) or the $1\!+\!1\!+\!1$ case ($m_x=m_y\not=m_u$ and
$m_x=m_y\not=m_d$) has double poles.  These can be dealt with as in
the quenched case: either take the limit $m_y\to m_x$ in
\eq{mass_an_terms} or (\ref{eq:mass_an_terms_21}), or return to
\eq{sigma_disc_combined} and perform the double pole integrals
directly.\footnote{If one chooses to perform the double pole
integrals directly, \eq{lagrange} is no longer valid, and a generalization
of this formula is needed.}

\section{Remarks and Conclusions}\label{sec:conc}

The most general result we have is for the $n=3$ partially quenched case ($1\!+\!1\!+\!1$)
with all valence and sea quark masses different,
\eq{mass_an_terms}.  Other interesting cases 
can be obtained from \eq{mass_an_terms} by taking appropriate mass limits. 
The results most relevant to current MILC simulations are those
with $m_u=m_d\equiv m_l$ (the $2\!+\!1$ case); these and other important
limits are presented explicitly in 
Sec.~\ref{sec:final_results}.
The result in the quenched case is given separately in
\eq{qu_an_terms}.

At this point, one can calculate any other desired quantity within
this framework. The calculation for the pion and kaon decay constants is 
straightforward; a description is now being prepared for publication\cite{AUBIN_BERNARD}. As in the
case here of the one loop pion mass, it is again simpler to
examine the partially quenched case, and from there all the necessary
results can be obtained. The next step will be the incorporation of
heavy quarks, so that we can examine the effects of staggered discretization errors
on heavy-light meson quantities. This requires an extension of
these ideas to incorporate the heavy quark symmetries within \schpt,
and is in progress.

The generalization of the Lee-Sharpe Lagrangian to multiple
flavors has shown that two additional parameters, $\delta'_V$ and
$\delta'_A$, appear in the one-loop chiral logarithms for the charged
meson masses. These parameters are not determined at tree level by existing 
lattice data for pion mass splittings, since they contribute only to
unmeasured disconnected tree graphs for flavor-neutral, taste-nonsinglet,
pions.  The new parameters are therefore unconstrained in current
chiral-logarithm fits to lattice results.  In contrast, the masses of
the charged pions of various tastes that appear in our final results
are not free parameters in the one-loop fits, since they are
determined at tree-level by lattice measurements.  Using tree-level
information, a fit of lattice data to \eq{mass_an_terms_21} would have
7 free parameters: $\mu$, $f$, $2L_8-L_5$, $2L_6-L_4$, $\delta'_V$, $\delta'_A$
and $C$.\footnote{One may choose to absorb $C$ into $\mu$, which will 
have $\cO(a^2)$ corrections in any case from  higher operators in 
the effective continuum action that have the same symmetries as the 
lowest order terms --- see Appendix.  However, this will change
the higher order corrections to \eq{mass_an_terms_21}.}
We remark  that existence of the parameters $\delta'_V$ and
$\delta'_A$ leads to the possibility of phase transition before the chiral
limit of the staggered theory is reached. This possibility is discussed
further in the Appendix; it does not appear to be realized in practice
for the strange quark mass at its physical value.

Despite the presence of additional parameters, well controlled  simultaneous
fits to partially quenched lattice results for $f_\pi$, $f_K$, $m^2_\pi/(2m_l)$
and $m^2_K/(m_l + m_s)$ at fixed lattice spacing appear possible \cite{MILC_FITS}.  
These should allow for highly accurate extrapolations to physical quark mass
and then to the continuum, as well as determinations of the Gasser-Leutwyler
parameters $L_i$.  It can help here to constrain, at least weakly, the
new chiral parameters.  One easy way to do this is to use a vacuum saturation
estimate of the matrix elements of the 4-quark taste-violating operators calculated
in perturbation theory \cite{MASON}.  More accurate lattice evaluations of
the matrix elements, or perhaps even direct lattice determinations of the $\delta'_V$ and
$\delta'_A$ by evaluation of disconnected pion propagators, may also be envisioned.

An alternative approach to the fitting of lattice data is also possible when
highly accurate data exists at more than one lattice spacing.  Here one can
extrapolate to the continuum at fixed quark mass and then fit the resulting
``continuum'' results to standard \chpt\ forms, \ie without taste violations.
This is the approach  taken in \cite{PRL}, and it works well.
Because of the nonanalytic dependence on the lattice spacing induced
by the chiral logarithms coming from pions of various tastes,  though,
there is a residual discretization error left in the data even after extrapolation
to the continuum. This error would go away if one worked very close to the continuum
limit, where ``very close'' here means $ka^2 \Lambda^2_{QCD} \ll m_\pi^2/\Lambda^2$,
with $k$ a constant that depends on the particular staggered action,
and $\Lambda$ is the chiral scale. For pions light enough for \chpt\ to be applicable,
however, this condition is very difficult to satisfy without further
improvement in the staggered action than is currently available.   
The \schpt\  formulas above will therefore remain crucial, at least in the near term,
for determining the systematic errors in the results.

\bigskip
\bigskip
\centerline{\bf ACKNOWLEDGMENTS}
\bigskip
We thank M.\ Golterman, G.\ P.\ Lepage,
S.\ Sharpe, and D.\ Toussaint for helpful discussions.
This work was partially supported by the U.S. Department of Energy
under grant number DE-FG02-91ER40628.

\bigskip
\bigskip
\centerline{\bf APPENDIX}
\bigskip
Here we write down the symmetries of the effective continuum action 
(``Symanzik action'') of the staggered lattice theory through $\cO(a^2)$,
and those of the corresponding chiral theory, \eq{final_L}.
We also briefly discuss the interesting
possibility of a transition of the staggered theory to an unusual phase.
We follow the notation and reasoning of Ref.~\cite{LEE_SHARPE} closely; the
discussion in this Appendix is {\it not} self-contained.

The symmetries of various terms in the $\cO(a^2)$ Symanzik action
are shown in Table \ref{tab:symm}, which is a 
generalization of 
Table 1 in Ref.~\cite{LEE_SHARPE} to the current $n$-flavor case.

\begin{table}
\renewcommand{\arraystretch}{1.4}
\begin{center}
\begin{tabular}{llll}
\hline
Term in action &&& [Flavor] $\times$ Rotation symmetry \\ \hline
$S_4$ ($m=0$)           &&& $[U(1)_{\rm VEC}\times SU(4n)_L\times SU(4n)_R] \times SO(4)$ \\
$S_4$ ($m\ne0$)         &&& $[(\;U(1)_{\rm vec}\times SU(4)_{\rm vec}\;)^n] \times SO(4)$ \\
\hline
$S_6^{\rm glue}$        &&& $[U(1)_{\rm VEC}\times SU(4n)_L\times SU(4n)_R] \times SW_4$ \\
$S_6^{\rm bilin}$ ($m=0$)&&& $[U(1)_{\rm VEC}\times SU(4n)_L\times SU(4n)_R] \times SW_4$ \\
$S_6^{\rm bilin}$ ($m\ne0$)&&& $[(\;U(1)_{\rm vec}\times SU(4)_{\rm vec}\;)^n] \times SW_4$ \\
$S_6^{\rm FF(A)}$       &&&
$[U(n)_\ell \times U(n)_r\times(\Gamma_4 \semitimes SO(4))] \times SO(4)$ \\
$S_6^{\rm FF(B)}$       &&&
$U(n)_\ell \times U(n)_r \times (\Gamma_4 \semitimes SW_{4,\rm diag})$ \\
\hline
$S_6 (m=0)$             &&&
$U(n)_\ell \times U(n)_r \times (\Gamma_4 \semitimes SW_{4,\rm diag})$ \\
$S_6 (m\ne 0)$          &&&
$(\;U(1)_{\rm vec}\;)^n \times \Gamma_4 \semitimes SW_{4,\rm diag}$ \\
\hline
\end{tabular}
\caption{The flavor 
and rotation symmetries respected
by various terms in the effective action.
Here ``flavor'' is used generically to include fermion number, true vector flavor, 
chiral, and  taste symmetries.
Almost all the notation is from Ref.~\cite{LEE_SHARPE}. 
The ``residual chiral group,'' $U(n)_\ell \times U(n)_r$ is
defined in the text. We have also added
the subscript ``vec'' for vector ($L+R$) symmetries, and
have included overall fermion number, $U(1)_{\rm VEC}$, as well individual flavor number
symmetries, $U(1)_{\rm vec}$.
There is no clear separation of flavor and rotation symmetries in the last three lines.
For simplicity in the $m\ne0$ cases, we assume that all quark masses are nonzero 
and different for different flavors.
}
\label{tab:symm}
\end{center}
\end{table}

The  ``residual chiral group,'' $U(n)_\ell \times U(n)_r$, which is a symmetry of
$S^{\rm FF(A)}_6$ and $S^{\rm FF(B)}_6$, is the extension to multiple flavors
of the residual $U(1)_{\rm vec}\times U(1)_A$ symmetry of a single staggered field.
Let $t_a$ be the $U(n)$ generators, and let $q$ be the complete (flavor $\otimes$ taste 
$\otimes$ spin) quark field, as in \eq{four-quark} but with flavor indices suppressed.
Then the residual chiral group is given by:

\begin{eqnarray}\label{eq:lr}
  \ell & : &\cases {q\to \exp\left[i\theta^a_\ell t_a 
    \left({\displaystyle \frac{1 - 
	\gamma_5\otimes\xi_5}{2}}\right)\right] q\ , \cr
  \bar q \to \bar q\; \exp\left[-i\theta^a_\ell t_a 
    \left({\displaystyle\frac{1 + \gamma_5\otimes\xi_5}{2}}\right) \right] ;}
  \nonumber\\*
  r&: & \cases {q\to \exp\left[i\theta^a_r t_a 
    \left({\displaystyle \frac{1 + 
	\gamma_5\otimes\xi_5}{2}}\right)\right] q\ ,\cr
    \bar q \to \bar q\; \exp\left[-i\theta^a_r t_a 
    \left(\displaystyle\frac{1 - \gamma_5\otimes\xi_5}{2} \right) \right] \ ;\cr}
\end{eqnarray}
where $\theta^a_\ell$ and $\theta^a_r$ ($a=1,2,\dots,n^2$) are the group parameters.
We use the notation $\ell$ and $r$, rather than the usual $L$ and $R$ for
chiral rotations, because these symmetries combine chiral spin with taste.
To study the effect of this symmetry on various terms in Table~\ref{tab:symm}, 
consider a flavor-singlet, ``odd'' bilinear, \ie a bilinear of the form
$ \bar{q}(\gamma_S\otimes\xi_T)q$, where there is an implicit sum
over flavor, and where
$\{\gamma_S\otimes\xi_T,\gamma_5\otimes\xi_5\}=0$.  
Then it is clear that any bilinear of this type is invariant
under the  residual chiral symmetry, \eq{lr}.
That the $S^{\rm FF(A)}$ and
$S^{\rm FF(B)}$ terms in the action are invariant under this 
symmetry now follows from the fact they can be written
as sums of products of such bilinears (see
discussion preceding  \eq{U}). 

Note that, even though the identity matrix in flavor is included
among the generators $t_a$ in \eq{lr}, the corresponding axial
symmetries ($\theta_\ell^a = -\theta_r^a$) are traceless in flavor 
$\otimes$ taste because of the presence of $\xi_5$.

For future purposes it is convenient to rewrite \eq{lr} to show
explicitly the action of the $\ell$ and $r$ symmetries on the
chiral fields.  Define 
\begin{eqnarray}\label{eq:LR}
& q_L   \equiv  \left({\displaystyle\frac{1-\gamma_5}{2}}\right)\; q\ , 
\qquad  & \bar q_L  = \bar q \left(\frac{1+\gamma_5}{2}\right)\nonumber \\*
& q_R \equiv  \left({\displaystyle\frac{1+\gamma_5}{2}}\right)\; q\ , 
\qquad  & \bar q_R  = \bar q \left(\frac{1-\gamma_5}{2}\right) \\*
& U_{\ell L}  \equiv \exp\left[i\theta^a_\ell t_a 
\left({\displaystyle \frac{1 + \xi_5}{2}}\right)\right] \ ,
\qquad & U_{\ell R}    \equiv \exp\left[i\theta^a_\ell t_a 
\left( \frac{1 - \xi_5}{2}\right)\right] \nonumber \\*
& U_{r L}   \equiv \exp\left[i\theta^a_r t_a 
\left({\displaystyle \frac{1 - \xi_5}{2}}\right)\right] \ ,
 \qquad & U_{r R}   \equiv \exp\left[i\theta^a_r t_a 
\left( \frac{1 + \xi_5}{2}\right)\right]\ .
\end{eqnarray}
Then 
\begin{eqnarray}\label{eq:lrLR}
\ell: & \quad q_L \to   U_{\ell L}\; q_L  \ ,
& \qquad \bar q_L \to \bar q_L\; U^\dagger_{\ell L} \nonumber \\*
& \quad q_R \to   U_{\ell R}\; q_R  \ ,
& \qquad \bar q_R \to \bar q_R\; U^\dagger_{\ell R} \nonumber \\*
r: & \quad q_L \to   U_{r L}\; q_L  \ ,
& \qquad \bar q_L \to \bar q_L\; U^\dagger_{r L} \nonumber \\*
& \quad q_R \to   U_{r R}\; q_R  \ ,
& \qquad \bar q_R \to \bar q_R\; U^\dagger_{r R} 
\end{eqnarray}

One now assumes that the $SU(4n)_L\times SU(4n)_R$ approximate symmetry 
(\ie the symmetry of $S_4$, the 4-dimensional terms in the action, at $m=0$) breaks
dynamically in the usual way down to $SU(4n)_{\rm vec}$.  
The kinetic energy term in the effective chiral Lagrangian then has
the complete $SU(4n)_L\times SU(4n)_R$ symmetry (realized nonlinearly).
Other terms in the Symanzik action are represented by additional terms
in the chiral Lagrangian with the corresponding symmetries.

A key insight of Lee and Sharpe is that the chiral representatives of
all terms in the action that violate the $SO(4)$
rotation symmetry must contain derivatives.
For example, the rotationally noninvariant
term $\sum_\mu \bar q(\gamma_\mu \otimes I)D^3_\mu q\;$ in $S^{\rm bilin}_6$ 
has a lowest chiral representive
$\sum_\mu Tr (\partial^2_\mu \Sigma \partial^2_\mu \Sigma^\dagger)$.
Chiral terms that are already $\cO(a^2)$ and also have derivatives will
be higher order than our $\cO(m,a^2)$ Lagrangian, \eq{final_L}.
Thus only $S_4(m=0)$, $S_4(m\not=0)$, and $S^{\rm FF(A)}_6$ contribute
to \eq{final_L}, giving the kinetic energy, mass term, and potential
$\cV$, respectively.  The symmetry group of the chiral Lagrangian is therefore
simply the intersection of the symmetry groups of the relevant three lines
from Table~\ref{tab:symm}:\footnote{We
are ignoring the discrete symmetries of parity and charge conjugation here.}
$[(\;U(1)_{\rm vec}\;)^n \times(\Gamma_4 \semitimes SO(4))] \times SO(4)$,
although by treating the mass and taste violating matrices as spurions,
one can work with the full
$[U(1)_{\rm VEC}\times SU(4n)_L\times SU(4n)_R] \times SO(4)$ group.

Under the 
 the residual chiral symmetry $U(n)_\ell\times U(n)_r$, \eq{lrLR}, the chiral
field $\Sigma$ transforms as
\begin{eqnarray}\label{eq:lrLRSigma}
\ell: & \quad \Sigma \to U_{\ell L}\; \Sigma\; U_{\ell R}^\dagger\ ,
& \qquad \Sigma^\dagger \to U_{\ell R}\;\Sigma^\dagger\; U_{\ell L}^\dagger \nonumber \\*
r: & \quad \Sigma \to U_{r L}\; \Sigma\; U_{r R}^\dagger\ ,
& \qquad \Sigma^\dagger \to U_{r R}\;\Sigma^\dagger\; U_{r L}^\dagger 
\end{eqnarray}
It is straightforward to check that the kinetic energy and potential terms in
the Lagrangian, \eq{final_L}, are invariant under this symmetry, which is
of course violated by the mass term.

Note that terms violating continuum rotational invariance can appear
for the first time at $\cO(m a^2)$, from the chiral representatives of
$S^{\rm FF(B)}_6$.  However, because the taste of the Goldstone pion transforms 
trivially under lattice rotations ($SW_{4,\rm diag}$), rotational violations
cannot affect it unless four derivatives are present.
Thus, for example, the Goldstone pion's continuum dispersion relation is violated
at  $\cO(m^2 a^2)$ by a term $a^2 \sum_\mu \partial^2_\mu \pi_5\; \partial^2_\mu \pi_5$
coming from the chiral reprentatives of the noninvariant terms in
$S^{\rm bilin}_6$ and $S_6^{\rm glue}$.

\begin{figure}
 \includegraphics[width=3in]{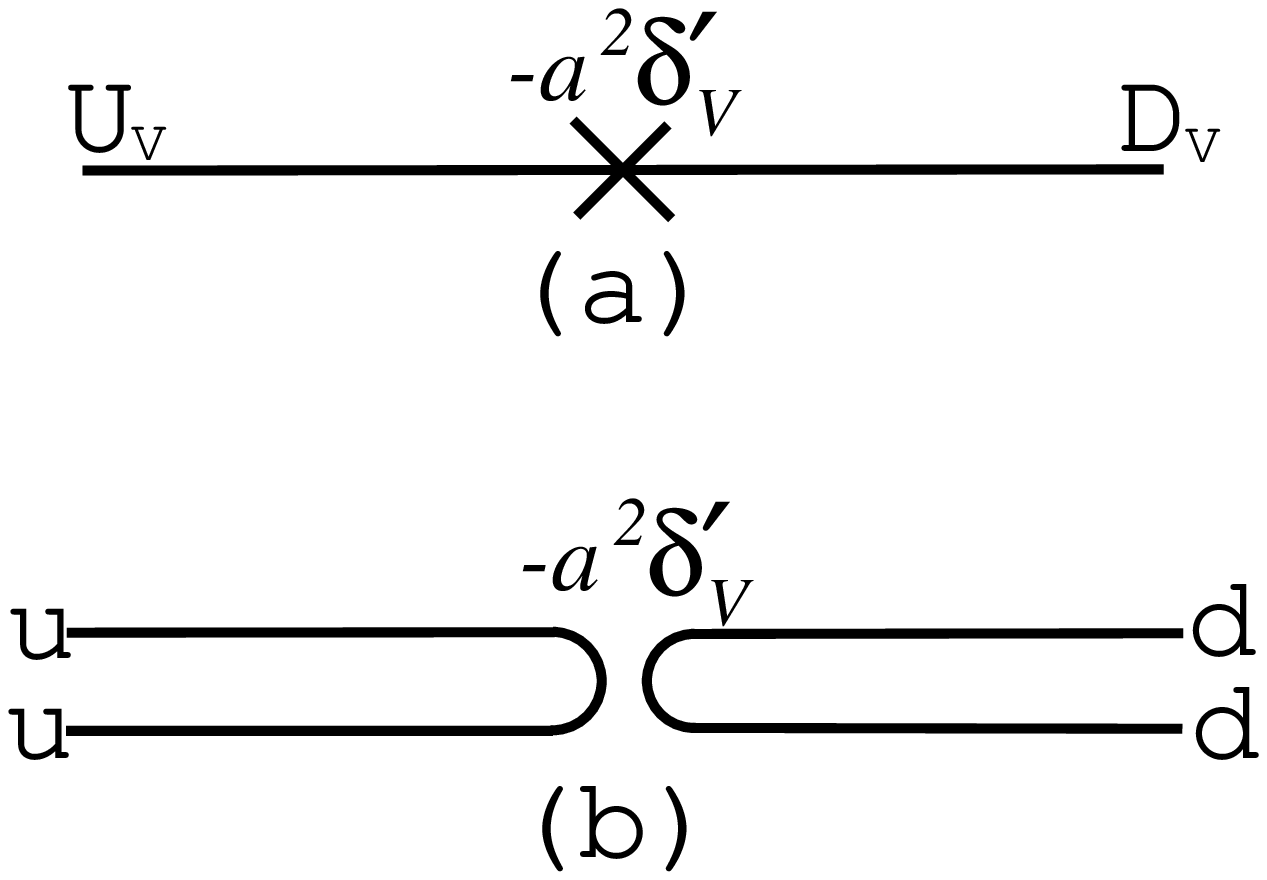} \caption{The two-point
  	mixing vertex (among taste vectors) coming from the new $\cU\,'$ term. 
	(a) corresponds to the chiral theory.  (b) shows the corresponding
  	quark level diagram.  We also have $U$-$S$ and $D$-$S$ mixings
  	and diagonal terms ($U$-$U$ \etc).  There are similar vertices
  	among the axial tastes (with $a^2\delta'_V \to a^2\delta'_A$), as 
	well as the singlet tastes (with $a^2\delta'_V \to 4m_0^2/3)$.}  
	\label{fig:2-pt_vertex}
\end{figure}

The $SO(4)$ part of the taste symmetry of the lowest order
chiral action guarantees that the approximate spectral degeneracies 
found in Ref.~\cite{LEE_SHARPE} persist in the $n$-flavor theory. This 
symmetry is ``accidental'' in the sense that it is not obeyed by the full
lattice action and will be violated at next order. Note that the
taste $SO(4)$, and in fact the accompanying discrete Clifford 
group $\Gamma_4$, appear only once, as an overall taste symmetry affecting
all flavors, and not as individual groups for each flavor separately.
This can be seen from the structure of the four-quark operators, \eq{four-quark}.
It is related to the fact that the symmetry of the underlying lattice theory
has just a single $\Gamma_4 \semitimes SW_{4,{\rm diag}}$ factor, which is generated
by single site translations and lattice rotations. It clearly cannot act
on different flavors separately since the gauge fields must also be translated 
and rotated.  We remark further that, if there were $\Gamma_4 \semitimes SO(4)$ 
taste symmetries for each flavor separately, they would forbid taste-nonsinglet
hairpins graphs like Fig.~\ref{fig:2-pt_vertex}.  

Lee and Sharpe have discussed the possibility of an unusual 
``Aoki-phase'' for staggered fermions that could
occur if the mass squared of one of the non-Goldstone pions vanished before
the chiral limit. However, since the splittings, $\Delta(\xi_B)$  in \eq{deltas}
are all positive for existing staggered actions, this scenario seems
unlikely to be realized in practice.  

The current work suggests another
possibility for an unusual phase: from \eq{eigenvalues_V}, if $\delta'_A$ or
$\delta'_V$ is negative and sufficiently large in magnitude compared to
the Goldstone masses and to
the corresponding splittings, $\Delta_A$ or $\Delta_V$ (\eq{deltas}),
then $m^2_{\eta_A}$ or $m^2_{\eta_V}$ could vanish before the chiral limit.
This possibility seems to us not as remote as the previous one, because
chiral logarithm fits \cite{MILC_FITS} to existing MILC data tend to give a negative
value for $\delta'_A$ that is comparable in magnitude to $\Delta_A$.  Taking
$m_u=m_d$, \eq{eigenvalues_V} implies that $m^2_{\eta_A}$ vanishes before
the chiral limit ($m_u=m_d=0$) is reached if
\begin{equation}\label{eq:phase_transition}
\delta'_A < \delta'_{A,\,{\rm crit}} \equiv  -4\Delta_A\; \frac{1 + a^2\Delta_A/m^2_{S_5}}
{2 + 3a^2\Delta_A/m^2_{S_5}}
\end{equation}
For the $s$-quark mass at its physical value, $m^2_{S_5} \approx (700 \MeV)^2$. On
the ``coarse'' ($a\approx 0.13$ fm) MILC lattices, $a^2\Delta_A \approx (275 \MeV)^2$.
This means that the transition could occur with a physical strange
quark mass only for $\delta'_A\; \ltwid -1.9 \Delta_A$, which does not appear to be
satisfied by the chiral fits. Further, as $a$ decreases, the fit values of $\delta'_A$
seem to move further from $\delta'_{A,\,{\rm crit}}$.
However, the transition appears considerably more likely to be realized in the unphysical
case where all three quark masses get small.  There $\delta'_{A,\,{\rm crit}} = -4\Delta_A/3$,
which is comparable to fit values of $\delta'_A$.    More study of this interesting
possibility is clearly warranted.

\vfill\eject

\begin{thebibliography}{20}
\expandafter\ifx\csname natexlab\endcsname\relax\def\natexlab#1{#1}\fi
\expandafter\ifx\csname bibnamefont\endcsname\relax
  \def\bibnamefont#1{#1}\fi
\expandafter\ifx\csname bibfnamefont\endcsname\relax
  \def\bibfnamefont#1{#1}\fi
\expandafter\ifx\csname citenamefont\endcsname\relax
  \def\citenamefont#1{#1}\fi
\expandafter\ifx\csname url\endcsname\relax
  \def\url#1{\texttt{#1}}\fi
\expandafter\ifx\csname urlprefix\endcsname\relax\def\urlprefix{URL }\fi
\providecommand{\bibinfo}[2]{#2}
\providecommand{\eprint}[2][]{\url{#2}}

\bibitem{CHIRAL_PANEL_2001}
C.\ Bernard, \et,
\npbps{106-107}, 199 (2002).

\bibitem{FB_LAT02}
C.\ Bernard, \et\  (The MILC collaboration),
\boston, hep-lat/0208041.

\bibitem{IMP_SCALING}
C.\ Bernard, \et\  (The MILC collaboration),
\prd{61}, 111502(R) (2000).

\bibitem{IMP_SCALING2}
C.\ Bernard, \et\  (The MILC collaboration),
\prd{62}, 034503 (2000).

\bibitem{MILC_SPECTRUM}
C.\ Bernard, \et\  (The MILC collaboration),
\prd{64}, 054506 (2001).

\bibitem{LEE_SHARPE}
W.\ Lee and S.\ Sharpe,
\prd{60}, 114503 (1999).

\bibitem{CHIRAL_FSB}
C.\ Bernard,
\prd{65}, 054031 (2001).

\bibitem{SHARPE_QCHPT}
S.\ Sharpe,
\prd{46}, 3146 (1992); Nucl.\ Phys.\ \textbf{B} (Proc.\
  Suppl.) \textbf{17}, 1 (1990).

\bibitem{LAT02}
C.\ Aubin \et\  (The MILC Collaboration),
 hep-lat/0209066.

\bibitem{SHARPE_SHORESH}
S.\ Sharpe and N. Shoresh
\prd{64}, 114510 (2001).

\bibitem{LEPAGE}
We thank G.\ P.\ Lepage for discussions on this point
and for sharing with us his unpublished notes on the relation between the
naive and staggered theories.

\bibitem{UNPHYSICAL}
S.\ Sharpe and N. Shoresh
\prd{62}, 094503 (2000).

\bibitem{CBMG_PQCHPT}
C.\ Bernard and M. Golterman,
\prd{49}, 486 (1994).

\bibitem{CBMG_QCHPT}
C.\ Bernard and M. Golterman,
\prd{46}, 853 (1992).

\bibitem{GASSER_LEUTWYLER}
J.\ Gasser and H. Leutwyler,
\npb{250}, 465, 1985.

\bibitem{LAGRANGE}
See, for instance, G.~Carrier, M.~Krook, C.~Pearson,
\textit{Functions of a Complex Variable}, 
McGraw-Hill (1966), page 70; or see Ref.~\protect{\cite{LEE_SHARPE}}.

\bibitem{AUBIN_BERNARD}
C.\ Aubin and C.\ Bernard, in preparation.

\bibitem{MILC_FITS}
The MILC collaboration, work in
progress.

\bibitem{MASON}
Q.\ Mason \et,
\boston, hep-lat/0209152.

\bibitem{PRL}
C.\ Davies \et\  (HPQCD, MILC, FERMILAB, and UKQCD collaborations),
hep-lat/0304004.

\end{thebibliography}
\end{document}